\documentclass[pre,twocolumn,nofootinbib]{revtex4-1}
\usepackage{amsmath,amssymb}
\usepackage[dvipdfmx]{graphicx}
\usepackage[dvipdfmx]{hyperref}
\usepackage{bm}
\usepackage{mathrsfs}
\usepackage{braket}
\usepackage{comment}
\usepackage{color}
\newcommand{\mJ}{\mathcal{J}}
\newcommand{\mH}{\mathcal{H}}

\newcommand{\mG}{\mathcal{G}}

\makeatletter

\begin{document}

\title{Theory of rigidity and numerical analysis of density of states of two-dimensional amorphous solids with dispersed frictional grains in the linear response regime}
\author{Daisuke Ishima$^{1}$, }
\altaffiliation[Present Address: ]{Simplex Holdings, Inc., 19F Toranomon Hills Mori Tower, 1-23-1 Toranomon, Minato-ku, Tokyo 105-6319, Japan; ishima.daisuke.m30@kyoto-u.ac.jp}
\author{Kuniyasu Saitoh$^{2}$, Michio Otsuki$^{3}$ and Hisao Hayakawa$^{1}$}
\affiliation{
$^{1}$Yukawa Institute for Theoretical Physics, Kyoto University, \\
Kitashirakawa-oiwake cho, Sakyo-ku, Kyoto 606-8502, Japan,\\
$^{2}$Department of Physics, Faculty of Science, Kyoto Sangyo University,\\
Motoyama, Kamigamo, Kita-ku, Kyoto 603-8555, Japan,\\
$^{3}$Graduate School of Engineering Science, Osaka University, Toyonaka, Osaka 560-8531, Japan
}
\date{\today}
\begin{abstract}
Using the Jacobian matrix, 
we obtain theoretical expression of rigidity and the density of states of two-dimensional amorphous solids consisting of frictional grains in the linear response to an infinitesimal strain, in which we ignore the dynamical friction caused by the slip processes of contact points.
The theoretical rigidity agrees with that obtained by molecular dynamics simulations.
We confirm that the rigidity is smoothly connected to the value in the frictionless limit.
For the density of states, we find that there are two modes in the density of states for sufficiently small $k_{T}/k_{N}$, which is the ratio of the tangential to normal stiffness.
Rotational modes exist at low frequencies or small eigenvalues, whereas translational modes exist at high frequencies or large eigenvalues.
The location of the rotational band shifts to the high-frequency region with an increase in $k_{T}/k_{N}$ and becomes indistinguishable from the translational band for large $k_{T}/k_{N}$.
\end{abstract}
\maketitle

\section{Introduction\label{intro}}

Amorphous materials consisting of dispersed grains such as powders, colloids, bubbles, and emulsions are ubiquitous in nature \cite{Jaeger96,Durian95,Wyart05,Baule18}.
These materials behave like liquids at low densities and exhibit solid-like mechanical responses above their jamming point \cite{Liu98}.
In systems consisting of frictionless grains, the rigidity changes continuously, but the coordination number of grains changes discontinuously at the jamming point as a function of density \cite{Durian95,Wyart05,Hern03}.
The critical behavior near the jamming point is of interest to physicists as a non-equilibrium phase transition \cite{Hern01,Trappe01,Zhang05,Majmudar07,Ciamarra11}.
Dispersed grains above the jamming point are fragile and exhibit softening and yielding transition under certain loads \cite{Nagamanasa14,Knowlton14,Kawasaki16,Leishangthem17,Bohy17,Ozawa18,Clark18,Boschan19,Singh20,Otsuki21}.

For amorphous solids consisting of frictionless grains, it is useful to analyze the dynamical matrix or the Hessian matrix,
which is defined as the second derivative of the potential of a collection of grains with respect to the displacements from their stable configuration \cite{Wyart05,WyarLettt05,Ellenbroek06, Lerner16, Gartner16, Bonfanti19, Baule18}.
For instance, the rigidity can be determined by eigenvalues and eigenvectors \cite{DeGiuli14, Mizuno16, Maloney04, Maloney06, Lemaitre06,Zaccone11}.
It has been reported that the minimum nonzero eigenvalue of the Hessian matrix decreases with increasing strain and eventually becomes negative, where an irreversible stress drop takes place~\cite{Maloney04, Maloney06, Manning11, Dasgupta12, Ebrahem20}.

It is known that amorphous solids have characteristic properties at low temperatures (e.g., thermal conductivity and specific heat) that are quite different from those of crystalline solids since a long-time ago \cite{Zeller71}. 
These days, we have recognized that amorphous solids consisting of dispersed grains exhibit unique elastic-plastic behavior as a mechanical response to an applied strain~\cite{Nicolas18}.
Because these properties are related to the density of states (DOS), there have been lots of studies on the DOS \cite{Hern03, Schirmacher07, Lerner16, Mizuno17, Wang19}.
The DOS for systems composed of anisotropic grains, such as ellipses, dimers, deformable grains, and grains with rough surfaces, have been studied with the aid of the Hessian matrix \cite{Zeravcic09,Mailman09,Schreck10,Yunker11,Schreck12, Shiraishi19, Papanikolaou13,Britoa18,Treado21,Ikeda20,Ikeda21}.
Because of the rotation of such anisotropic grains, there exists a rotational band in the DOS that is distinguishable from the translational band \cite{Mailman09,Yunker11,Schreck12,Britoa18,Ikeda20,Ikeda21}.

Even for systems of spherical grains that cannot be free from inter-particle friction, similar results are expected as a result of grain rotations.
However, few studies have reported the existence of rotational bands in the DOS.
Because the frictional force between the grains depends on the contact history, it cannot be expressed as a conservative force.
Therefore, stability analysis for frictional grains based on the Hessian cannot be used.
Nevertheless, the Hessian analysis using an effective potential for frictional grains has been performed \cite{Somfai07, Henkes10}.
Recently, Liu et al. suggested that the Hessian analysis with another effective potential can be used even if slip processes exist \cite{Liu21}.
The previous studies \cite{Somfai07,Henkes10} reported that friction between grains causes a continuous change in the functional form of the DOS from that of frictionless systems.
However, there are only a few reports on whether an isolated band in the DOS originating from friction between grains is visible at lower frequencies.

Recently, Chattoraj et al. discussed the stability of the grain configuration under strain using the Jacobian matrix of frictional grains \cite{Chattoraj19}.
They performed eigenvalue analysis under athermal quasi-static shear processes and determined the existence of oscillatory instability originating from inter-particle friction at a certain strain \cite{Chattoraj19, Chattoraj19E, Charan20}.
However, they did not discuss the rigidity or the DOS.

The theoretical determination of the rigidity of amorphous solids consisting of frictional grains is important for controlling amorphous solids.
However, we do not know how to determine the rigidity from the Jacobian for the frictional grains.

The purpose of this study is to clarify the role of mutual friction between grains in terms of the rigidity and DOS.
We focus on the response to an infinitesimal strain from a stable configuration of grains without any strain to obtain tangible results.
In this study, we assume that there is no slip between grains because of an infinitesimal strain, and we then deal with friction as static friction.

The remainder of this paper is organized as follows. In the next section, we introduce the numerical method.
In Sec. \ref{Sec3}, we introduce the Jacobian.
Section \ref{Sec4} consists of Sec \ref{ThG}, which deals with the theoretical prediction of rigidity in the linear response regime, and Sec. \ref{vDOS}, which deals with the DOS. 
In the final section, we summarize the results of our study and discuss future work.
The appendix consists of seven sections.
In Appendix \ref{AppIni}, we summarize the method for preparing a stable grain configuration before applying shear.
In Appendix \ref{AppVerlet}, we explain the implementation of the numerical integration method in the proposed system.
In Appendix \ref{AppC}, we summarize some properties of the Jacobian.
In Appendix \ref{AppD}, we present the explicit expressions of the Jacobian.
In Appendix \ref{appRatt}, we investigate the effects of rattlers. 
In Appendix \ref{JacobianHertz}, we write down the explicit results of Jacobian. 
In Appendix \ref{AppE}, we derive the theoretical prediction of rigidity using the Jacobian.
In Appendix \ref{AppF}, we introduce the DOS using the Hessian analysis.
In Appendix \ref{SystemSize}, we investigate the system size dependence of the DOS.
In Appendix \ref{DensityDep}, we study the density dependence of the DOS.

\section{Numerical model \label{Sec2}}

\begin{figure}[htbp]
\centering
\includegraphics[width=9cm]{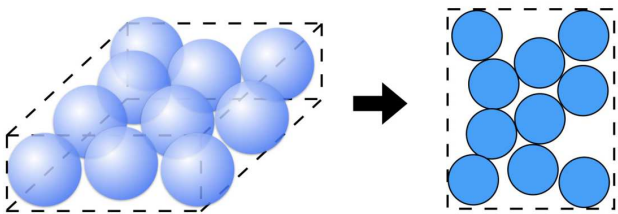}
\caption{
Schematics of our system.
}
\label{map}
\end{figure}

Our system contains $N$ frictional spherical particles embedded in a monolayer configuration.
We treat this system as a two-dimensional system (see Fig. \ref{map}).
To prevent the system from crystallizing \cite{Luding01}, we prepare an equal number of particles with diameters $d$ and $d/1.4$.
We assume that the mass of particle $i$ is proportional to $d_i^2$, where $d_i$ is the diameter of $i$-th particle.
We introduce $m$ as the mass of a particle with diameter $d$.
In this study, $x_{i}, y_{i}$, and $\theta_{i}$ denote $x, y$ coordinates, and the rotational angle of the $i$-th particle, respectively.
We introduce the generalized coordinates of the $i$-th particle $\bm{q}_{i}:=(\bm{r}_{i}^{\textrm{T}},\ell_{i})^{\textrm{T}}$ with $\bm{q}_i:=(x_{i},y_{i})^{\textrm{T}}$ and $\ell_i:=d_{i}\theta_{i}/2$, where the superscript $\textrm{T}$ denotes the transposition.

Let the force, and $z$-component of the torque acting on the $i$-th particle be $\bm{F}_{i}:=(F_{i}^{x},F_{i}^{y})^{\textrm{T}}$ and $T_{i}$, respectively.
Then, the equations of motion of $i$-th particle are expressed as
\begin{align}
m_{i}\frac{d^{2}\bm{r}_{i}}{dt^{2}} &=  \bm{F}_{i}, \label{tra}\\
I_{i}\frac{d^{2}\theta_{i}}{dt^{2}} &= T_{i} \label{rot}
\end{align}
with  the mass $m_i$ and the momentum of inertia $I_i:=m_i d_i^2/8$ of $i$-th particle.
In a system without volume forces such as gravity, we can write
\begin{align}
\bm{F}_{i}&=\sum_{j\neq i} \bm{f}_{ij}, \label{Fi}\\
T_{i}&=\sum_{j\neq i}T_{ij},
\end{align}
where $\bm{f}_{ij}$ and $T_{ij}$ are the force and $z$-component of the torque acting on the $i$-th particle from the $j$-th particle, respectively.
Here, $T_{ij}$ is given by
\begin{align}
T_{ij}=-\frac{d_{i}}{2}( {n}_{ij}^{x}{f}_{ij}^{y} - {n}_{ij}^{y}{f}_{ij}^{x} ),
\label{Torque}
\end{align}
where we have introduced the normal unit vector between $i$ and $j$ particles as $\bm{n}_{ij}:=\bm{r}_{ij}/|\bm{r}_{ij}|:=(\bm{r}_{i}-\bm{r}_{j})/|\bm{r}_{i}-\bm{r}_{j}|$.
Here, $n_{ij}^{\zeta}$ and $f_{ij}^{\zeta}$ refer to $\zeta$-components of $\bm{n}_{ij}$ and $\bm{f}_{ij}$, respectively.
Note that $\zeta$ expresses $\zeta=x$ or $y$ throughout this study.
The force $\bm{f}_{ij}$ can be divided into normal $\bm{f}_{N,ij}$ and tangential $\bm{f}_{T,ij}$ parts as
\begin{align}
\bm{f}_{ij}=(\bm{f}_{N,ij}+\bm{f}_{T,ij} )\Theta(d_{ij}/2-|\bm{r}_{ij}|)
\label{fijSum}
,
\end{align}
where $d_{ij}:=d_i+d_j$ and $\Theta(x)$ is Heaviside's step function, taking $\Theta(x)=1$ for $x>0$ and $\Theta(x)=0$ otherwise.
We model the repulsive force between the contacted particles $i$ and $j$ as the Hertzian force
in addition to the dissipative force proportional to the relative velocity with a damping constant $\eta_{D}$
\cite{Liu21} as follows:
\begin{align}
\bm{f}_{N,ij}:&=k_{N}\xi_{N,ij}^{3/2}\bm{n}_{ij} - \eta_{D}\bm{v}_{N,ij}
\label{fijn}
,\\
\bm{f}_{T,ij}:&=k_{T}\xi_{N,ij}^{1/2}\xi_{T,ij}\bm{t}_{ij} - \eta_{D}\bm{v}_{T,ij}
\label{fijt}
,
\end{align}
where $k_N$ and $k_T$ are the stiffness parameters of normal and tangential contacts, respectively.
For the normal compression length and its velocity are denoted as 
$\xi_{N,ij}:=d_{ij}/2-|\bm{r}_{ij}|$ and $\bm{v}_{N,ij}:=(\dot{\bm{r}}_{ij}\cdot\bm{n}_{ij})\bm{n}_{ij},$ respectively.
For the tangential deformation, with the aid of
$
\bm{u}_{ij}:=(n_{ij}^{y}, -n_{ij}^{x})^{\textrm{T}}$, the tangential velocity $\bm{v}_{T,ij}$ is defined as 
$\bm{v}_{T,ij}:=\dot{\bm{r}}_{ij}-\bm{v}_{N,ij}+\bm{u}_{ij}(d_{i}\omega_{i}+d_{j}\omega_{j})/2$, where we have introduced
\begin{align}
\bm{\xi}_{T,ij}:=\int_{C_{ij}}dt\bm{v}_{T,ij}
 -\left[\left( \int_{C_{ij}}dt\bm{v}_{T,ij} \right)\cdot\bm{n}_{ij}\right]\bm{n}_{ij} , 
\label{xiT}
\end{align}
with $\xi_{T,ij}:=|\bm{\xi}_{T,ij}|$
and 
$
\bm{t}_{ij}:=-\bm{\xi}_{T,ij}/\xi_{T,ij}
$.
Here, $\dot{\bm{A}}:=d\bm{A}/dt$ and $\int_{C_{ij}}dt$ is the integration over the duration time of contact between $i$ and $j$ particles.
Although the dissipative force between grains interacting with the Hertzian force is proportional to the product of the relative velocity and $\xi_{N,ij}^{1/2}$ \cite{Kuwabara87,Brilliantov96,Morgado97},
we adopt simple dissipative forces as in Eqs. \eqref{fijn} and \eqref{fijt} because we are not interested in the relaxation dynamics.
We note that Eqs. \eqref{fijn} and \eqref{fijt} assume the Hertzian contact force for the static repulsion of contacting spheres,
but all calculations in this study are those for two-dimensional systems.
Here, we do not consider the effects of slips in the tangential equation of motion.
This treatment can be justified if we restrict our interest in the linear response regime to a stable configuration of particles without any strain.
This situation corresponds to frictional grains with infinitely large dynamical friction constant, in which the friction is only characterized by static friction.
Therefore, our analysis does not apply to systems with finite strain \cite{Otsuki17}, where the effect of slip is important.

To generate a stable configuration of frictional particles, we prepare a stable configuration of frictionless particles in a square box of linear size $L$ using a fast inertial relaxation engine (FIRE) \cite{Bitzek06}.
Subsequently, we turn on the tangential force using Eqs. \eqref{tra} and \eqref{rot} to achieve a stable configuration in the force-balanced (FB) state for frictional particle\footnote{
For simplicity, we prepare the configuration before applying shear for frictionless particles at first, and then considered the friction between particles.
If we prepare a configuration before applying shear by compressing frictional particles,
we confirm that the configuration had an oscillatory instability that resulted from the appearance of a pair of imaginary eigenvalues of the Jacobian divided by mass matrix: $\lambda'=\lambda'_{r}\pm i\lambda'_{i}$ \cite{Chattoraj19,Chattoraj19E,Charan20},
where $\lambda',\lambda'_{r}$, and $\lambda'_{i}$ are the complex, real, and imaginary eigenvalues of $M^{-1}\mJ$, respectively.
Here, $\mathcal{J}$ is the Jacobian defined in Eq. \eqref{Jacobian},  and $M$ is the mass matrix whose explicit form is given by $M=\left[\begin{matrix}M_{1}& \\ &\ddots\\ & & M_{N} \end{matrix}\right]$, where $M_{i}:=\left[\begin{matrix}m_{i} & & \\ &m_{i}&\\ & & 4I_{i}/d_{i}^{2} \end{matrix}\right]$.
Because the linearized equation of motion is expressed as
$d^{2}q_{i}^{\alpha}/dt^{2}=-\sum_{k,\kappa}\sum_{j,\beta}(M_{ik}^{\alpha\kappa})^{-1}\mJ_{kj}^{\kappa\beta} q_{j}^{\beta}$,
there are four fundamental solutions $\bm{q}\propto e^{i\omega_{n}' t}$,
where $i\omega_{n}'$ consists of $i\omega'_{\pm1}=\omega'_{i}\pm i\omega'_{r}$ and $i\omega'_{\pm2}=-\omega'_{i}\pm i\omega'_{r}$.
Here, $\omega'_r$ and $\omega'_i$ satisfy the relation $\omega'_{r}\pm i\omega'_{i}=\sqrt{\lambda'_{r}\pm i\lambda'_{i}}$.
Thus, to avoid the oscillatory instability of the configuration before applying shear,
we adopt the protocol of creating the configuration with frictionless particles, and then let the system relax by adding static friction between particles.
} (see Appendix \ref{AppIni} for details).
Here, the FB state satisfies the FB conditions $|F_{i}^{\zeta}|=0$ and $|T_{i}|=0$ for arbitrary particles.
Note that we set $\theta_{i}=0$ when the tangential force is turned on.

We impose the Lees-Edwards boundary condition \cite{Lees72,Evans08},
where the direction parallel to the shear strain is $x$-direction.
After applying a step strain $\Delta\gamma$ to all particles, $x$-coordinate of the position of the $i$-th particle is shifted by an affine displacement $\Delta x_{i}(\Delta\gamma):=\Delta\gamma y_{i}^{\textrm{FB}}(0)$,
where the superscript FB denotes the FB state.
The system is then relaxed to the FB state by the contact forces between the particles expressed in Eqs. \eqref{fijn} and \eqref{fijt}.
Here, $\zeta$-components of translational $\Delta\mathring r_{i}^{\zeta}(\Delta\gamma)$ and rotational $\Delta\mathring\ell_{i}(\Delta\gamma)$ nonaffine displacements of the $i$-th particle after the relaxation process are, respectively, defined as
\begin{align}
\Delta\mathring r_{i}^{\zeta}(\Delta\gamma):&= r_{i}^{\textrm{FB},\zeta}(\Delta\gamma)-r_{i}^{\textrm{FB},\zeta}(0)-\delta_{\zeta x}\Delta\gamma y_{i}^{\textrm{FB}}(0), \label{AffTra}\\
\Delta\mathring \ell_{i}:&= \ell_{i}^{\textrm{FB}}(\Delta\gamma)-\ell_{i}^{\textrm{FB}}(0). \label{AffRot}
\end{align}
Using Eqs. \eqref{AffTra} and \eqref{AffRot}, we introduce the rate of nonaffine displacements as:
\begin{align}
\frac{d\mathring r_{i}^{\zeta}}{d\gamma}:&=
\lim_{\Delta\gamma\to0}\frac{\Delta\mathring r_{i}^{\zeta}(\Delta\gamma)}{\Delta\gamma} \nonumber \\
&= \lim_{\Delta\gamma\to0}\frac{r_{i}^{\textrm{FB},\zeta}(\Delta\gamma)-r_{i}^{\textrm{FB},\zeta}(0)}{\Delta\gamma}-\delta_{\zeta x} y_{i}^{\textrm{FB}}(0), \label{drdG}
\\
\frac{d\mathring \ell_{i}}{d\gamma}:&=
\lim_{\Delta\gamma\to0}\frac{\Delta\mathring \ell_{i}(\Delta\gamma)}{\Delta\gamma} \nonumber \\
&= \lim_{\Delta\gamma\to0}\frac{\ell_{i}^{\textrm{FB}}(\Delta\gamma)-\ell_{i}^{\textrm{FB}}(0)}{\Delta\gamma}
. \label{dldG}
\end{align}

Our system is characterized by the generalized coordinate $\bm{q}(\gamma):=(\bm{q}_{1}^{\textrm{T}}(\gamma), \bm{q}_{2}^{\textrm{T}}(\gamma), \cdots, \bm{q}_{N}^{\textrm{T}}(\gamma))^{\textrm{T}}$.
The configuration in the FB state at strain $\gamma$ is denoted as $\bm{q}^{\textrm{FB}}(\gamma):=((\bm{q}_{1}^{\textrm{FB}}(\gamma))^{\textrm{T}}, (\bm{q}_{2}^{\textrm{FB}}(\gamma))^{\textrm{T}},\cdots,(\bm{q}_{N}^{\textrm{FB}}(\gamma))^{\textrm{T}})^{\textrm{T}}$.
The shear stress $\sigma_{xy}(\gamma)$ at $\bm{q}^{\textrm{FB}}(\gamma)$ for one sample is given by:
\begin{align}
\sigma_{xy}(\bm{q}^{\textrm{FB}}(\gamma))=
-\frac{1}{L^{2}} \sum_{i}\sum_{j>i} f_{ij}^{x}(\bm{q}^{\textrm{FB}}(\gamma)) r_{ij}^{y}(\bm{q}^{\textrm{FB}}(\gamma))
\label{Sigma}
.
\end{align}
The rigidity $g$ in the linear response regime for one sample is defined as
\begin{align}
g&:= \left. \frac{d\sigma_{xy}(\bm{q}(\gamma))}{d\gamma} \right|_{\bm{q}(\gamma)=\bm{q}^{\textrm{FB}}(0)}
\label{g}
,
\end{align}
where the differentiation on the right-hand side (RHS) of Eq. \eqref{g} is defined as follows:
\begin{align}
&\left. \frac{d\sigma_{xy}(\bm{q}(\gamma))}{d\gamma} \right|_{\bm{q}(\gamma)=\bm{q}^{\textrm{FB}}(0)}
\nonumber \\
&\quad\quad\quad:=\lim_{\Delta \gamma\to 0}
\frac{\sigma_{xy}(\bm{q}^{\textrm{FB}}(\Delta\gamma)) - \sigma_{xy}(\bm{q}^{\textrm{FB}}(0))}{\Delta\gamma}
\label{Gnm}
.
\end{align}
In the numerical calculation, we use a non-zero but sufficiently small $\Delta \gamma$ for the evaluation of $g$.
Then, the rigidity $G$ in the linear response regime is defined as
\begin{align}
G&:=\Braket{ g }
\label{Gdef}
,
\end{align}
where $\langle \cdot \rangle$ is the ensemble average.

For the numerical FB condition, we use the condition  $|\tilde F_{i}^{\alpha}|<F_{\textrm{Th}}$ for arbitrary $i$, where $F_{\textrm{Th}}$ is the threshold force for the simulation and $\tilde{\bm{F}}_{i}$ is the generalized force, defined as $\tilde{\bm{F}}_{i}:=(\tilde F_{i}^{x}, \tilde F_{i}^{y}, \tilde F_{i}^{\ell})^{\textrm{T}}:=(F_{i}^{x}, F_{i}^{y}, 2T_{i}/d_{i})^{\textrm{T}}$.

In our simulation, we adopt $\eta_{D}=(mk_{N})^{1/2}d^{1/4}$ and $F_{\textrm{Th}}=1.0\times10^{-14}k_{N}d^{3/2}$.
The control parameters are the ratio of the tangential to normal stiffness $k_{T}/k_{N}$ and projected area fraction to two-dimensional space $\phi$.
The operating ranges of $k_{T}/k_{N}$ and $\phi$ are $0.0$ to $10.0$ and $0.83$ to $0.90$, respectively.
In this study, we mainly present the results for $N=4096$ and $\Delta\gamma=1.0\times10^{-6}$ with the ensemble averages of $10$ samples for each $k_{T}/k_{N}$ and $\phi$.
Some results are obtained with $N=1024$, $\Delta\gamma=1.0\times10^{-6}$, and five samples for each $k_{T}/k_{N}$ and $\phi$.
We verify that the results are independent of the choice of $\Delta\gamma$ for $1.0\times10^{-6}\leq\Delta\gamma\leq1.0\times10^{-4}$.
We ignore the effect of dissipation between particles because the velocity of each particle is sufficiently small for infinitesimal agitation from the FB state.
The time step of the simulation, $\Delta t$, was set to $\Delta t=1.0\times10^{-2}t_{0}$, and numerical integration was performed using the velocity Verlet method (see  Appendix \ref{AppVerlet}), where $t_{0}:=(m/k_{N})^{1/2}d^{1/4}$.
To obtain eigenvalues and eigenvectors of the Jacobian matrix, which will be introduced in detail in the next section, 
we have used the LAPACK, which provides a template library for linear algebra.

\begin{figure}[htbp]
\centering
\includegraphics[width=9cm]{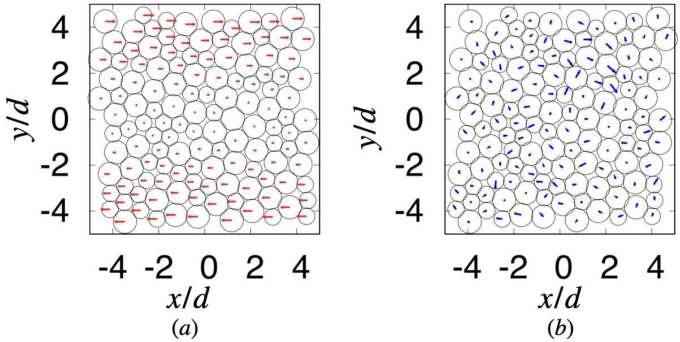}
\caption{
Plots of (a) affine displacements and (b) nonaffine displacements of particles with $\Delta\gamma=1.0\times10^{-6}$.
Here, the magnitudes of the vectors are multiplied by (a) $0.1\Delta\gamma^{-1}$ and (b) $1.3$ for the visualization.
These figures are based on numerical results for $N=128$.
}
\label{NVec}
\end{figure}

Figure \ref{NVec} (a) shows an example of the affine displacements of particles,
where the displacements exist only in the shear direction,
and Fig. \ref{NVec} (b) shows the nonaffine displacements.

\section{Theoretical Analysis \label{Sec3}}

In this section, we introduce the Jacobian, the DOS, and theoretical expressions of the linear rigidity.
Here, we omit the effects of $\dot{\bm{q}}$ because the dissipative term proportional to $\dot{\bm{q}}$ vanishes under quasistatic shear.

\subsection{Jacobian and the DOS for frictional particles \label{Sec3A}}

In frictional systems, the stability of the system and DOS at $\bm{q}^{\textrm{FB}}(\gamma)$ are analyzed using the Jacobian ($\mJ$) defined as \cite{Chattoraj19}:
\begin{align}
\mJ_{ij}^{\alpha\beta}:=-\left. \frac{\partial \tilde F_{i}^{\alpha}(\bm{q}(\gamma))}{\partial q_{j}^{\beta}} \right|_{\bm{q}(\gamma)=\bm{q}^{\textrm{FB}}(0)},
\label{Jacobian}
\end{align}
where $\alpha$ and $\beta$ are any of $x,y$ and $\ell$, while $i$ and $j$ express particle indices.
Therefore, the Jacobian matrix, which is a $3N\times3N$ matrix, can be written as
\begin{align}
\mJ=
\left[
\begin{matrix}
\mJ_{11} & \cdots & \mJ_{1i} & \cdots & \mJ_{1j} & \cdots & \mJ_{1N}\\
\vdots & \ddots & \vdots &   & \vdots &  &\vdots \\
\mJ_{i1} & \cdots & \mJ_{ii} & \cdots & \mJ_{ij} & \cdots & \mJ_{iN}\\
\vdots &  & \vdots & \ddots & \vdots &  &\vdots \\
\mJ_{j1} & \cdots & \mJ_{ji} & \cdots & \mJ_{jj} & \cdots & \mJ_{jN}\\
\vdots &  & \vdots &   & \vdots & \ddots &\vdots \\
\mJ_{N1} & \cdots & \mJ_{Ni} & \cdots & \mJ_{Nj} & \cdots & \mJ_{NN}
\end{matrix}
\right]
,
\label{defJ}
\end{align}
where $\mJ_{ij}$ is a $3\times3$ submatrix of the Jacobian $\mJ$ for a pair of particles $i$ and $j$:
\begin{align}
\mJ_{ij}=
\left[
\begin{matrix}
\mJ_{ij}^{xx} & \mJ_{ij}^{xy} & \mJ_{ij}^{x\ell} \\
\mJ_{ij}^{yx} & \mJ_{ij}^{yy} & \mJ_{ij}^{y\ell} \\
\mJ_{ij}^{\ell x} & \mJ_{ij}^{\ell y} & \mJ_{ij}^{\ell\ell}
\end{matrix}
\right]
\label{33J}
.
\end{align}
See Appendices \ref{AppC} and \ref{AppD} for detailed properties of the Jacobian.
The right and left eigenvalue equations of $\mJ$ are, respectively, given by
\begin{align}
\mJ\ket{R_{n}}&= \lambda_{n}\ket{R_{n}}, \label{REieq}\\
\bra{L_{n}}\mJ&=\lambda_{n}\bra{L_{n}}, \label{LEieq}
\end{align}
where $\ket{R_{n}}$ and $\bra{L_{n}}$ are the right and left eigenvectors corresponding to $\lambda_{n}$, respectively.
Here, $\lambda_{n}$ is the $n$-th eigenvalue of $\mJ$.
Note that $\ket{R_n}$ and $\bra{L_n}$ satisfy the orthonormal relation $\braket{L_m|R_n}=\delta_{mn}$ with normalization $\braket{R_n|R_n}=\braket{L_n|L_n}=1$, if all eigenstates are non-degenerate.
Here, the inner products for the right and left eigenvectors are defined as $\braket{R_{n}|R_{n}}=\sum_{i=1}^{N}\sum_{\alpha=x,y,\ell}|R_{n,i}^{\alpha}|^{2}$ and $\braket{L_{n}|L_{n}}=\sum_{i=1}^{N}\sum_{\alpha=x,y,\ell}|L_{n,i}^{\alpha}|^{2}$, respectively.
In the presence of friction, the eigenvalue $\lambda_n$ is generally a complex number, but if we restrict our interest to infinitesimal distortions from stable configurations without shear strain, $\lambda_n$ becomes real and can be expressed as $\lambda_n=\omega_n^2$.
The DOS is the distribution function of the eigenvalues, defined as:
\begin{align}
D(\omega):=\frac{1}{3N}\sum_{n}\nolimits' \langle \delta(\omega-\omega_n)\rangle , 
\label{defDOS}
\end{align}
where $\sum_{n}\nolimits'$ on RHS of Eq. \eqref{defDOS} expresses that the summation excludes the contribution of rattlers (see Appendix \ref{appRatt} for the details of rattlers). 
Using the force decomposition, the Jacobian can also be divided into
\begin{align}
\mJ_{ij}^{\alpha\beta}=\mJ_{N,ij}^{\alpha\beta}+\mJ_{T,ij}^{\alpha\beta}
\label{sepNT}
,
\end{align}
where
\begin{align}
\mJ_{N,ij}^{\alpha\beta}:&=\left. \frac{\partial \tilde f_{N,ij}^{\alpha}(\bm{q}(\gamma))}{\partial q_{j}^{\beta}} \right|_{\bm{q}(\gamma)=\bm{q}^{\textrm{FB}}(\gamma)},\\
\mJ_{T,ij}^{\alpha\beta}:&=\left. \frac{\partial \tilde f_{T,ij}^{\alpha}(\bm{q}(\gamma))}{\partial q_{j}^{\beta}} \right|_{\bm{q}(\gamma)=\bm{q}^{\textrm{FB}}(\gamma)}
\end{align}
for $i\neq j$ and
\begin{align}
\mJ_{N,ij}^{\alpha\beta}:&=\sum_{(i,k)}\left. \frac{\partial \tilde f_{N,ik}^{\alpha}(\bm{q}(\gamma))}{\partial q_{i}^{\beta}} \right|_{\bm{q}(\gamma)=\bm{q}^{\textrm{FB}}(\gamma)},\\
\mJ_{T,ij}^{\alpha\beta}:&=\sum_{(i,k)}\left. \frac{\partial \tilde f_{T,ik}^{\alpha}(\bm{q}(\gamma))}{\partial q_{i}^{\beta}} \right|_{\bm{q}(\gamma)=\bm{q}^{\textrm{FB}}(\gamma)}
\end{align}
for $i=j$.
Here, we have introduced $\tilde{\bm{f}}_{N,ij}:=(\tilde f_{N,ij}^{x}, \tilde f_{N,ij}^{y}, \tilde f_{N,ij}^{\ell} )^{\textrm{T}}:=(f_{N,ij}^{x}, f_{N,ij}^{y}, 0)^{\textrm{T}}$ and $\tilde{\bm{f}}_{T,ij}:=(\tilde f_{T,ij}^{x}, \tilde f_{T,ij}^{y}, \tilde f_{T,ij}^{\ell} )^{\textrm{T}}:=(f_{T,ij}^{x}, f_{T,ij}^{y}, 2T_{ij}/d_{i})^{\textrm{T}}$,
where $f_{N,ij}^{\zeta}$ and $f_{T,ij}^{\zeta}$ are $\zeta$-component of $\bm{f}_{N,ij}$ and $\bm{f}_{T,ij}$, respectively.
Note that $\sum_{(i,j)}$ denotes the summation for contacted particles of the $i$-th particle.
The explicit expressions of $\mJ_{N,ij}^{\alpha\beta}$ and $\mJ_{T,ij}^{\alpha\beta}$ are presented in Appendix \ref{JacobianHertz}

\subsection{Expressions of the linear rigidity via eigenmodes \label{Sec3B}}

Let us introduce $\ket{\tilde F(\bm{q}(\gamma))}$ as
\begin{align}
\ket{\tilde F(\bm{q}(\gamma))}:=[ \tilde{\bm{F}}_{1}^{\textrm{T}}(\bm{q}(\gamma)), \tilde{\bm{F}}_{2}^{\textrm{T}}(\bm{q}(\gamma)), \cdots, \tilde{\bm{F}}_{N}^{\textrm{T}}(\bm{q}(\gamma)) ]^{\textrm{T}}.
\end{align}
Because the forces acting on the particles are balanced in the FB state,
$\ket{\tilde{F}(\bm{q}(\gamma))}|_{\bm{q}(\gamma)=\bm{q}^{\textrm{FB}}(\gamma)}$ satisfies
\begin{align}
\left. \ket{\tilde F(\bm{q}(\gamma))} \right|_{ \bm{q}(\gamma)=\bm{q}^{\textrm{FB}}(\gamma) }=\ket{0},
\end{align}
where $\ket{0}$ is the ket vector containing $0$ for all components.
The stable configuration in the FB state satisfies
\begin{align}
\left. \frac{d\ket{\tilde F(\bm{q}(\gamma))}}{d\gamma} \right|_{\bm{q}(\gamma)=\bm{q}^{\textrm{FB}}(0)}
=\ket{0}
\label{dFdG}
,
\end{align}
where
\begin{align}
&\left. \frac{d\ket{\tilde F(\bm{q}(\gamma))}}{d\gamma} \right|_{\bm{q}(\gamma)=\bm{q}^{\textrm{FB}}(0)}
\nonumber \\
&\quad\quad\quad:=\lim_{\Delta \gamma\to 0}
\frac{\Ket{\tilde{F}(\bm{q}^{\textrm{FB}}(\Delta\gamma))} - \Ket{\tilde{F}(\bm{q}^{\textrm{FB}}(0))} }{\Delta\gamma}
\label{dFdG0}
.
\end{align}
Introducing
\begin{align}
\Ket{\frac{d\mathring q}{d\gamma}}
:=
\left[
\frac{d\mathring r_{1}^{x}}{d\gamma},
\frac{d\mathring r_{1}^{y}}{d\gamma},
\frac{d\mathring \ell_{1}}{d\gamma},
\cdots,
\frac{d\mathring r_{N}^{x}}{d\gamma},
\frac{d\mathring r_{N}^{y}}{d\gamma},
\frac{d\mathring \ell_{N}}{d\gamma}
\right]^{\textrm{T}}
,
\label{NonAff}
\end{align}
the left-hand side (LHS) of Eq.\eqref{dFdG} can be rewritten as:
 \begin{align}
\left. \frac{d\ket{\tilde F(\bm{q}(\gamma))}}{d\gamma} \right|_{\bm{q}(\gamma)=\bm{q}^{\textrm{FB}}(0)} =-\ket{\Xi} +\tilde\mJ\Ket{\frac{d\mathring q}{d\gamma}}
\label{dFdG2}
,
\end{align}
where we have used Eqs. \eqref{drdG} and \eqref{dldG} (see Appendix \ref{AppE1}).
The first and second terms on RHS of Eq. \eqref{dFdG2} represent
the strain derivatives of the forces for the contributions from the affine and nonaffine displacements, respectively.
The explicit form of $\ket{\Xi}$ is given by:
\begin{align}
\ket{\Xi}:=
\left[
\begin{matrix}
\sum_{j\neq 1}\mJ_{N,j1}^{xx}r_{1j}^{y}\\
\sum_{j\neq 1}\mJ_{N,j1}^{xy}r_{1j}^{y}\\
\sum_{j\neq 1}\mJ_{N,j1}^{x\ell}r_{1j}^{y}\\
\vdots\\
\sum_{j\neq N}\mJ_{N,jN}^{xx}r_{Nj}^{y}\\
\sum_{j\neq N}\mJ_{N,jN}^{xy}r_{Nj}^{y}\\
\sum_{j\neq N}\mJ_{N,jN}^{x\ell}r_{Nj}^{y}
\end{matrix}
\right]
.
\end{align}
Note that the tangential displacements do not contribute to $\ket{\Xi}$.
This is because the affine displacements are applied to our system instantaneously as a step strain; thus, the integral interval of the tangential displacement during the affine deformation is zero.
We have used $\tilde\mJ$ in Eq. \eqref{dFdG2} defined as
\begin{align}
\tilde\mJ_{ii}^{\alpha\beta}
&:=
\left\{
\begin{matrix}
-\mJ_{ii}^{\ell x} & (\alpha=\ell,\ \beta=x) \\
-\mJ_{ii}^{\ell y} & (\alpha=\ell,\ \beta=y) \\
\mJ_{ii}^{\alpha\beta} & (\textrm{otherwise})
\end{matrix}
\right.
\label{tildeJ1}
\end{align}
and
\begin{align}
\tilde\mJ_{ij}^{\alpha\beta}
&:=
\left\{
\begin{matrix}
-\mJ_{ij}^{x\ell} & (\alpha=x,\ \beta=\ell) \\
-\mJ_{ij}^{y\ell} & (\alpha=y,\ \beta=\ell) \\
\mJ_{ij}^{\alpha\beta} & (\textrm{otherwise})
\end{matrix}
\right.
\label{tildeJ2}
\end{align}
for $i\neq j$.

Expanding the nonaffine displacements by the eigenfunctions of $\tilde\mJ$ and using the fact that the LHS in Eq. \eqref{dFdG2} is zero, we obtain
\begin{align}
\Ket{\frac{d\mathring q}{d\gamma}}=\sum_{n}\nolimits' \frac{\braket{\tilde L_{n}|\Xi}}{\tilde{\lambda}_{n}}\ket{\tilde R_{n}},
\label{eqEiExp}
\end{align}
where $\tilde{\lambda}_{n},\bra{\tilde L_{n}}$, and $\ket{\tilde R_{n}}$ are the $n$-th eigenvalue of $\tilde\mJ$, and the left and right eigenvectors corresponding to $\tilde{\lambda}_n$, respectively.
Note that $\ket{\tilde R_n}$ and $\bra{\tilde L_n}$ satisfy the orthonormal relation $\braket{\tilde L_m|\tilde R_n}=\delta_{mn}$, if all eigenstates are non-degenerate.
See Appendix \ref{AppE1} for the derivation of Eq. \eqref{eqEiExp}.

The rigidity in the linear response regime under infinitesimal strain $\Delta\gamma$ is decomposed into two parts:
\begin{align}
G:=G_{\textrm{A}}+G_{\textrm{NA}} ,
\label{expG}
\end{align}
where $G_{\textrm{A}}$ and $G_{\textrm{NA}}$ are the rigidities corresponding to the affine and nonaffine displacements, respectively.
With the aid of Eqs. \eqref{Sigma}, \eqref{Gdef}, and \eqref{tildeJ2} the expressions of $G_{\textrm{A}}$ and $G_{\textrm{NA}}$ are, respectively, given by (see Appendix \ref{AppC2})
\begin{align}
G_{\textrm{A}}:&=
\frac{1}{2L^{2}}
\Braket{
\sum_{i,j(i\neq j)}
(r_{ij}^{y})^{2}\mJ_{N,ji}^{xx}
}
\label{GAff}
, \\
G_{\textrm{NA}}:&=
\frac{1}{2L^{2}}
\Braket{
\sum_{i,j(i\neq j)}
\Biggl[
\sum_{\zeta=x,y}r_{ij}^{y}\tilde\mJ_{ij}^{x\zeta}\frac{d\mathring r_{ij}^{\zeta}}{d\gamma}
-r_{ij}^{y}\tilde\mJ_{ij}^{x\ell} \frac{d\mathring \ell_{ij}}{d\gamma}
\Biggl]
}
,
\label{GNAff}
\end{align}
where we have introduced
\begin{align}
\frac{d\mathring r_{ij}^{\zeta}}{d\gamma}:&=\frac{d\mathring r_{i}^{\zeta}}{d\gamma} - \frac{d\mathring r_{j}^{\zeta}}{d\gamma}, \\
\frac{d\mathring \ell_{ij}}{d\gamma}:&=\frac{d\mathring \ell_{i}}{d\gamma} + \frac{d\mathring \ell_{j}}{d\gamma}.
\end{align}

Substituting Eq. \eqref{eqEiExp} into Eq. \eqref{GNAff}, $G_{\textrm{NA}}$ can be rewritten as follows:
\begin{align}
G_{\textrm{NA}}&=
-
\frac{1}{L^{2}}
\left\langle
\sum_{n}\nolimits' \frac{\braket{\tilde L_{n}|\Xi}\braket{\Theta|\tilde R_{n}}}{\tilde\lambda_{n}}
\right\rangle
\label{GNAffv2}
,
\end{align}
where we have introduced
\begin{widetext}
\begin{align}
\bra{\Theta}:=
\left[
\begin{matrix}
\sum_{j\neq 1 }r_{1j}^{y}\tilde{\mJ}_{j1}^{xx} ,
\sum_{j\neq 1 }r_{1j}^{y}\tilde{\mJ}_{j1}^{xy} ,
\sum_{j\neq 1 }r_{1j}^{y}\tilde{\mJ}_{j1}^{x\ell} ,
\cdots ,
\sum_{j\neq N }r_{Nj}^{y}\tilde{\mJ}_{jN}^{xx} ,
\sum_{j\neq N }r_{Nj}^{y}\tilde{\mJ}_{jN}^{xy} ,
\sum_{j\neq N }r_{Nj}^{y}\tilde{\mJ}_{jN}^{x\ell}
\end{matrix}
\right].
\end{align}
\end{widetext}
The affine rigidity can be also expressed as
\begin{align}
G_{\textrm{A}}= \frac{1}{L^2} \left\langle \braket{Y|\Xi} \right\rangle ,
\end{align}
where
\begin{align}
\bra{Y}:=\left[ r_{1j}^{y}, 0, 0, r_{2j}^{y},0, 0, \cdots, r_{Nj}^{y}, 0, 0 \right]
.
\end{align}

\section{Results \label{Sec4}}

In this section, we present the results of eigenvalue analysis and rigidity based on the formulation explained in the previous section.
In Sec. \ref{ThG}, rigidity is determined using the eigenmodes of the Jacobian.
Section \ref{vDOS} clarifies the effects of translational and rotational motions on the DOS.

\subsection{Theoretical evaluation of $G$\label{ThG}}

\begin{figure}[htbp]
\centering
\includegraphics[width=6cm]{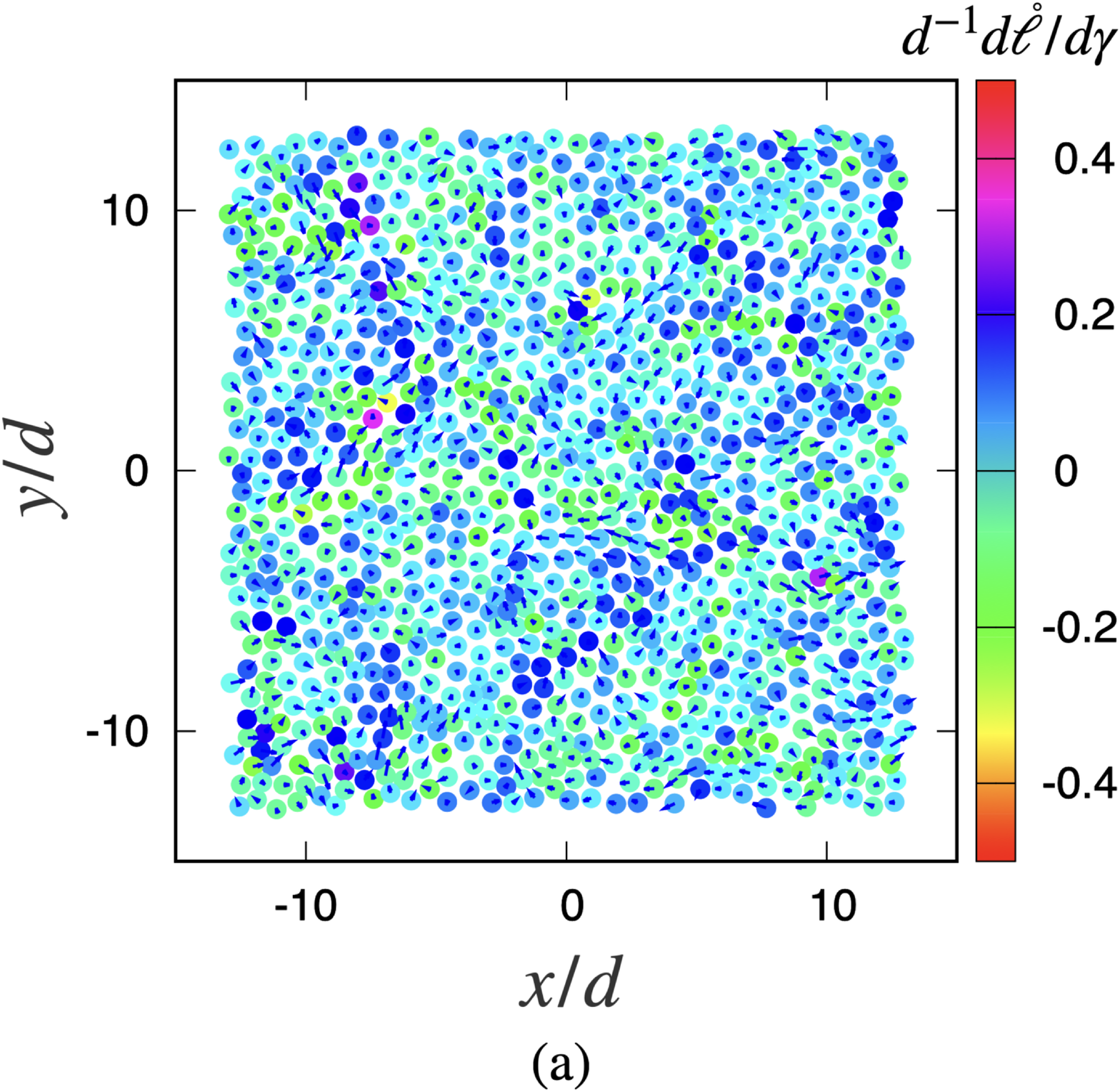}
\includegraphics[width=6cm]{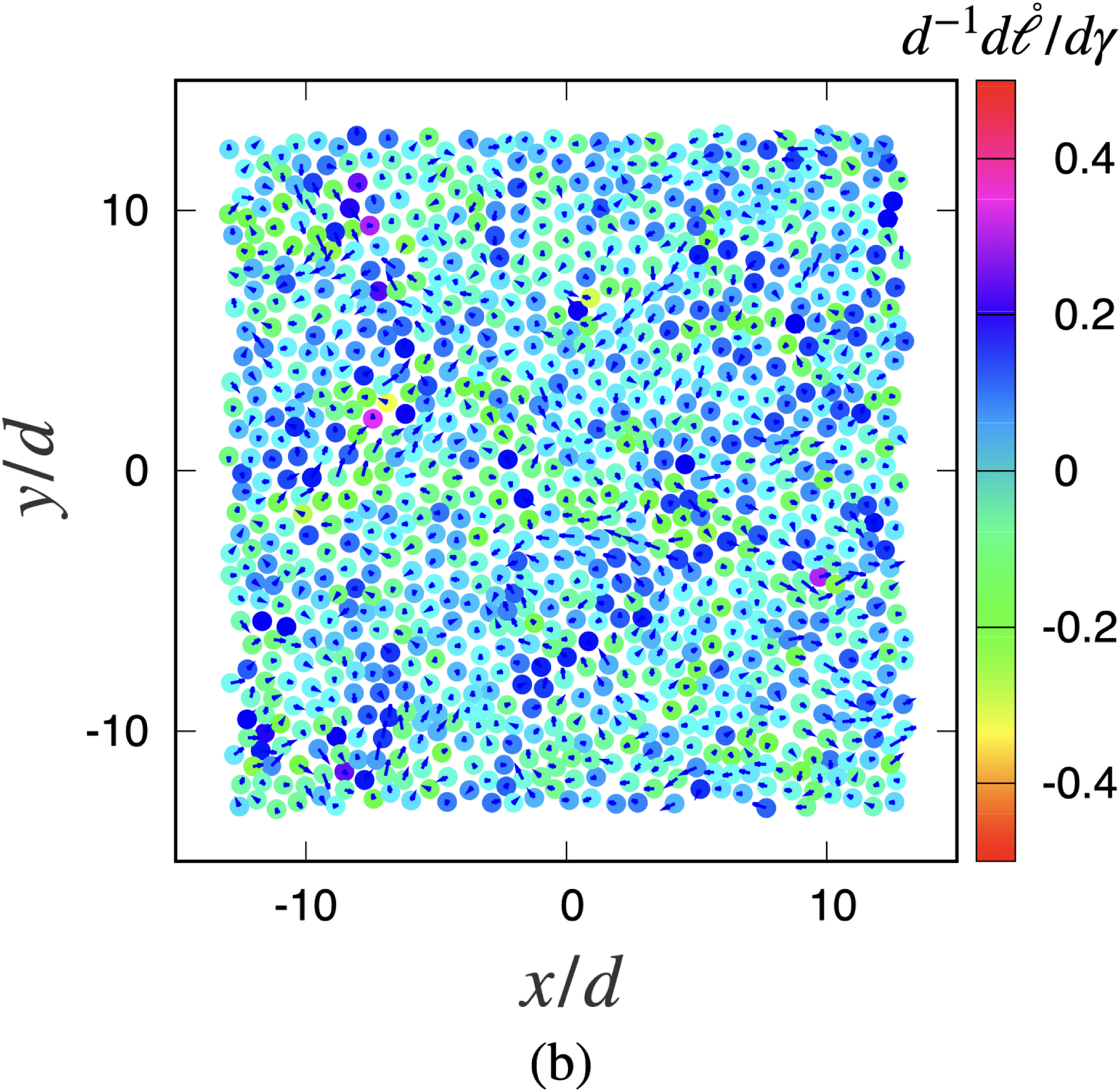}
\includegraphics[width=8cm]{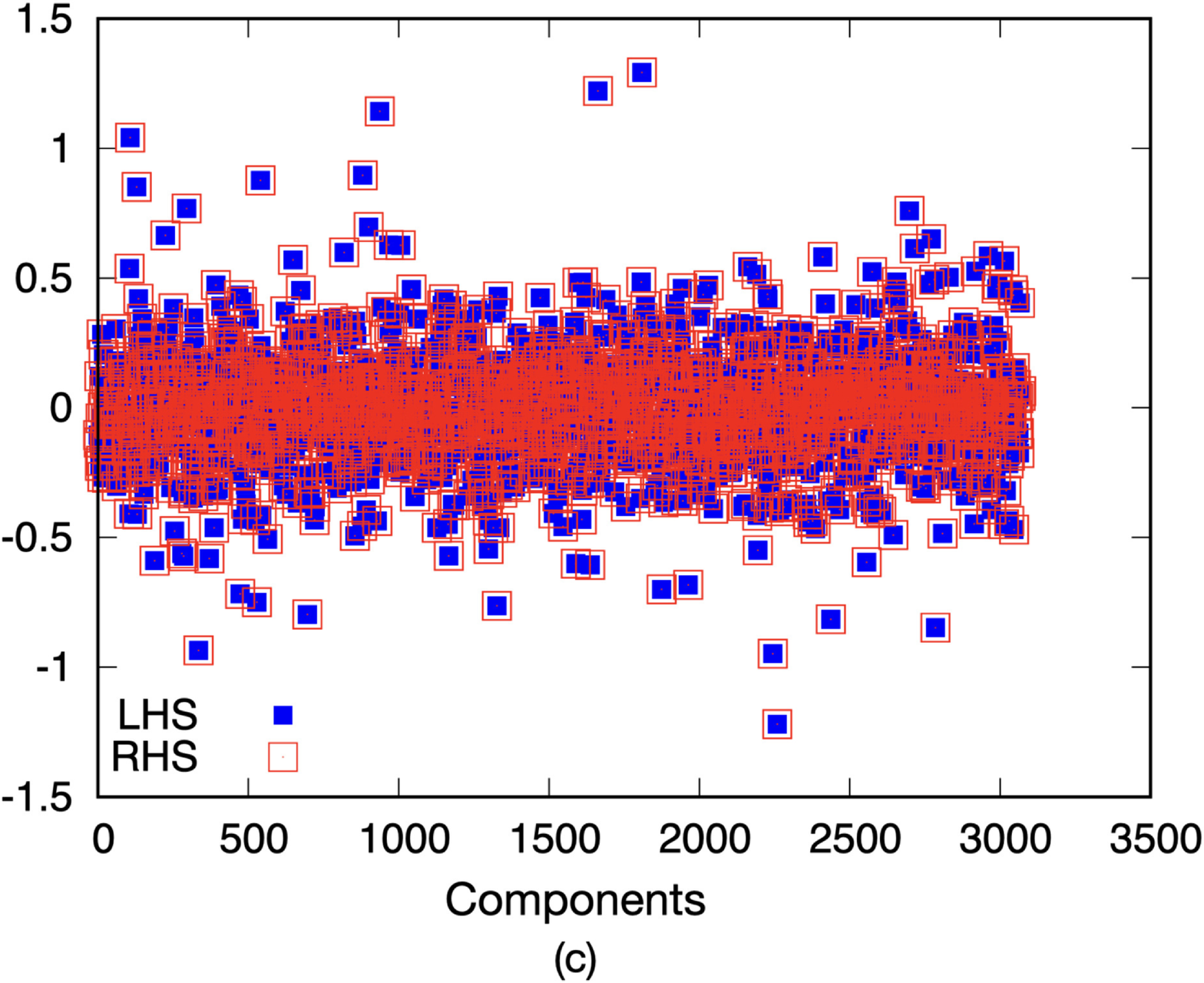}
\caption{
Plots of nonaffine displacements on (a) RHS of Eq. \eqref{eqEiExp} obtained by the eigenvalue analysis and (b) LHS of Eq. \eqref{eqEiExp}.
The vector and color of each particle correspond to $x,y,$ and $\ell$-components of the eigenvector of the particle, respectively.
Here, the magnitude of the vectors is magnified by $1.3$ times for visualization.
(c) Plots of the RHS (open symbols) and LHS (filled symbols) obtained by the simulation of Eq. \eqref{eqEiExp} for each component whose order follows Eq. \eqref{NonAff}.
These figures are based on numerical results for $N=1024$.
}
\label{figNA}
\end{figure}

In this subsection, the validity of the theoretical rigidity presented in the previous section is demonstrated.
For this purpose, at first, we examine the validity of Eq. \eqref{eqEiExp}, obtained by the eigenfunction expansion of the nonaffine displacements for RHS and by the simulation for LHS.
Figures \ref{figNA} (a) and (b) illustrate the nonaffine displacements on LHS and RHS of Eq. \eqref{eqEiExp}, respectively.
In Figs. \ref{figNA} (a) and (b), $(x,y)$ and $\ell$-components of $d\mathring{\bm{q}}_{i}/d\gamma$ at $\bm{r}_i$ are represented by vectors and colors, respectively.
Figure \ref{figNA} (c) shows the RHS and LHS of Eq. \eqref{eqEiExp} against the components of the vectors whose orders follow Eq. \eqref{NonAff}, that is, the local order of the component follows $x, y,$ and $\ell$ by fixing the particle number, and we align the components from the first particle to the $N$-th particle without omitting modes with extremely small and zero eigenvalues.
Figure \ref{figNA} shows that the expression in Eq. \eqref{eqEiExp} correctly reproduces the simulation results.

\begin{figure}[htbp]
\centering
\includegraphics[width=8cm]{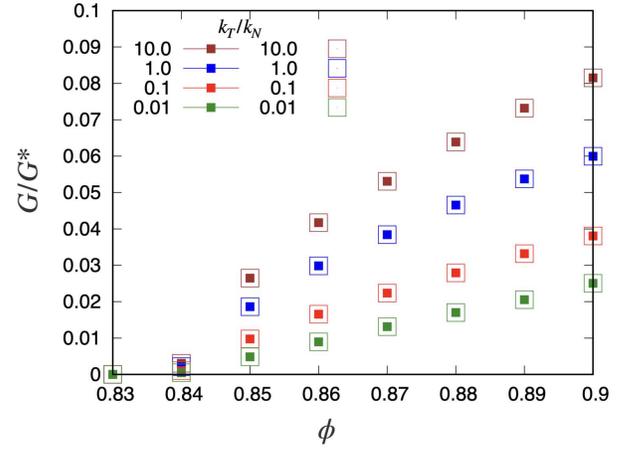}
\caption{
Plots of theoretical (Eq. \eqref{expG}: open symbols) and numerical (Eq. \eqref{Gnm}: filled symbols) $G$ against $\phi$ for various $k_T/k_N$.
The figure obtained by the numerical results for $N=1024$.
}
\label{figG}
\end{figure}

The dimensionless rigidity obtained from Eqs. \eqref{expG},\eqref{GAff} and \eqref{GNAffv2} with the aid of $G^*:=k_Nd^{1/2}$ is shown in Fig. \ref{figG}.
This indicates the quantitative agreement between the theoretical and numerical values.

Therefore, the rigidity in the linear response regime can be determined completely using the Jacobian analysis.
On the contrary to the previous studies \cite{Hern03,Olsson11,Otsuki11}, we should note that $G$ is not proportional to $\phi-\phi_{c}$ for a large $k_{T}/k_{N}$, where $\phi_{c}$ is the critical fraction of jamming transition for frictional grains.

\begin{figure}[htbp]
\centering
\includegraphics[width=8cm]{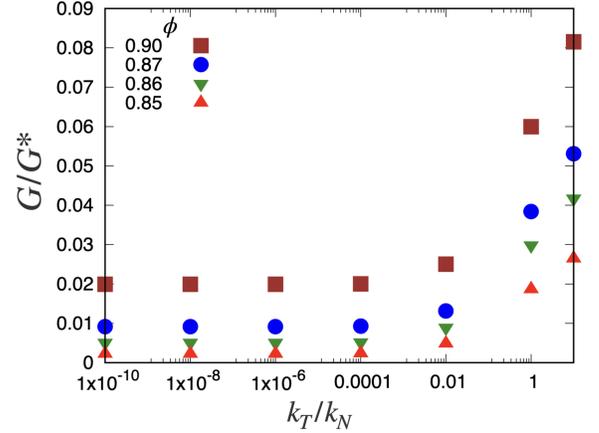}
\caption{
Plots of numerical $G$ against $k_{T}/k_{N}$ for various $\phi$.
The figure is obtained by the numerical results for $N=4096$.
}
\label{Gkt}
\end{figure}

As is expected, the rigidity $G$ depends on $k_{T}/k_{N}$ only little for $k_{T}/k_{N}\leq1.0\times10^{-4}$ (see Fig. \ref{Gkt}), while $G$ depends on $k_T/k_N$ for $k_T/k_N>10^{-4}$.
We have confirmed that $G$ smoothly approaches the frictionless value in the limit $k_T\to 0$ in contrary to Refs. \cite{Otsuki17, Ishima20}.
Here, $G$ cannot be expressed as a factorization for large $k_T/k_N$
\footnote{
We have confirmed that the factorization $G(\phi,k_{T}/k_{N})=G_{0}(\phi)\mG(k_{T}/k_{N})$ is not held,
where $G_{0}(\phi)$ is the rigidity of frictionless system.
}.
When we consider the effect of the dynamical friction, that is, slips between particles, the rigidity is discontinuously changed in the frictionless limit \cite{Otsuki17,Ishima20}.
However, the rigidity continuously changes with $k_{T}/k_{N}$ in our system and is smoothly connected to that of frictionless systems ($k_{T}=0$).
Because our system can be regarded as having infinitely large static and dynamical frictional constants,
there is no slip between the grains.
Therefore, it might be natural for $G$ to continuously change the limit for $k_{T}/k_{N}\to0$ in our system.
In future work, we will consider the effects of slips, which are important for real frictional grains.

\begin{figure}[htbp]
\centering
\includegraphics[width=8cm]{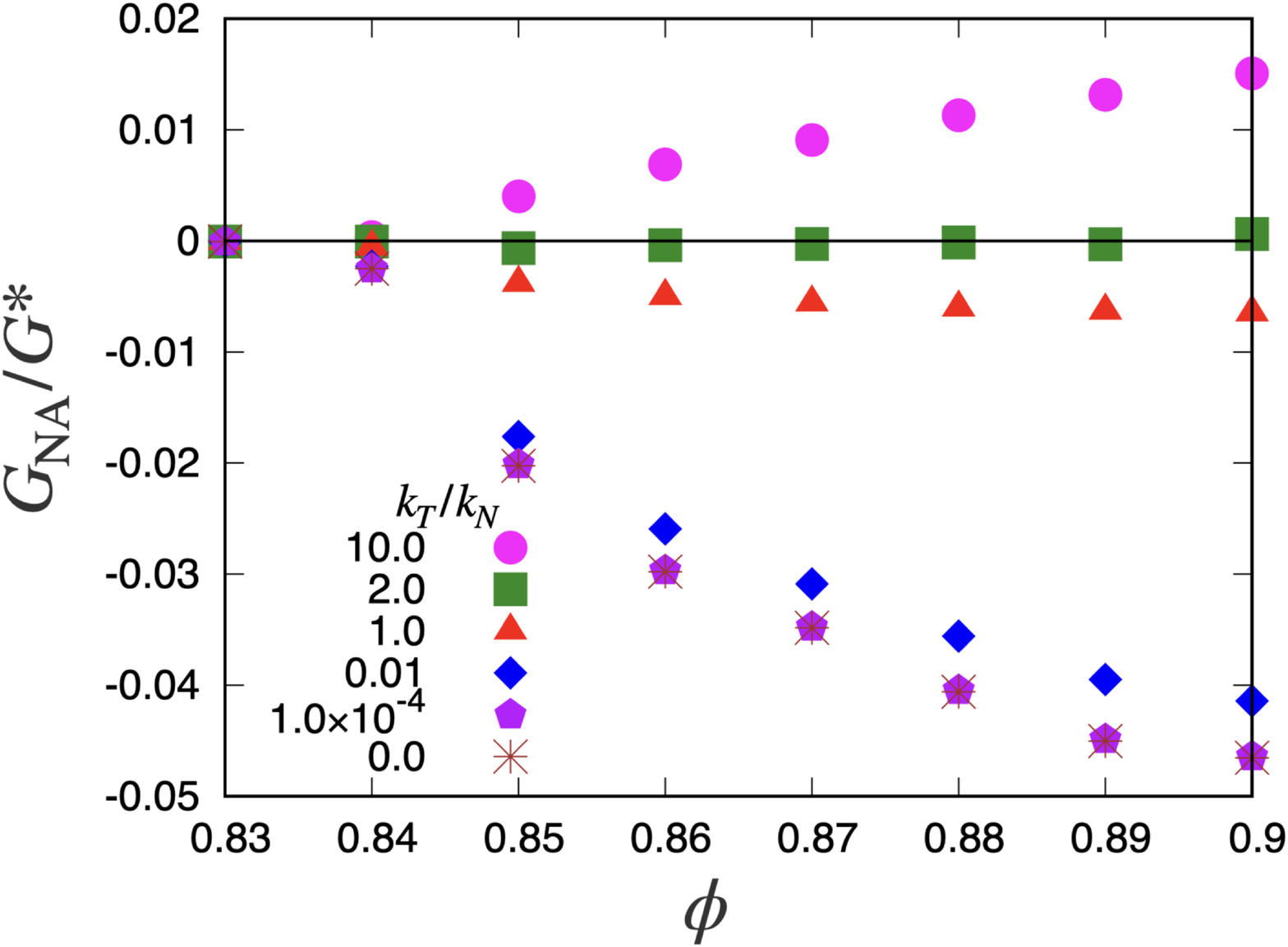}
\caption{
Plots of $G_{\textrm{NA}}$ against $\phi$ for various $k_T/k_N$.
The figure is obtained by the numerical results for $N=1024$.
}
\label{GA}
\end{figure}

To clarify the contributions of nonaffine deformations to the rigidity, we plot $G_{\textrm{NA}}$ defined in Eq. \eqref{GNAff} against $\phi$ in Fig. \ref{GA}, in which $G_{\textrm{NA}}$ becomes large as $\phi$ increases.
Remarkably, $G_{\textrm{NA}}$ is positive for $k_T/k_N>2.0$, $G_{\textrm{NA}}\approx 0$ at $k_T/k_N=2.0$, and $G_{\textrm{NA}}$ is negative for $k_T/k_N<2.0$.
The positive $G_{\textrm{NA}}$ for a large $k_T/k_N$ is counterintuitive, in which $G$ increases from $G_{\textrm{A}}$ even when the system is relaxed to the FB state.
In the future, we must clarify the origin of this counterintuitive $G_{\textrm{NA}}$.
We note that the negative $G_{\textrm{NA}}$ for a small $k_T/k_N$ can be understood by the relaxation process to look for a FB configuration after applying affine deformations to the system.

\subsection{Analysis of eigenvalues and eigenvectors \label{vDOS}}

\begin{figure}[htbp]
\centering
\includegraphics[width=9cm]{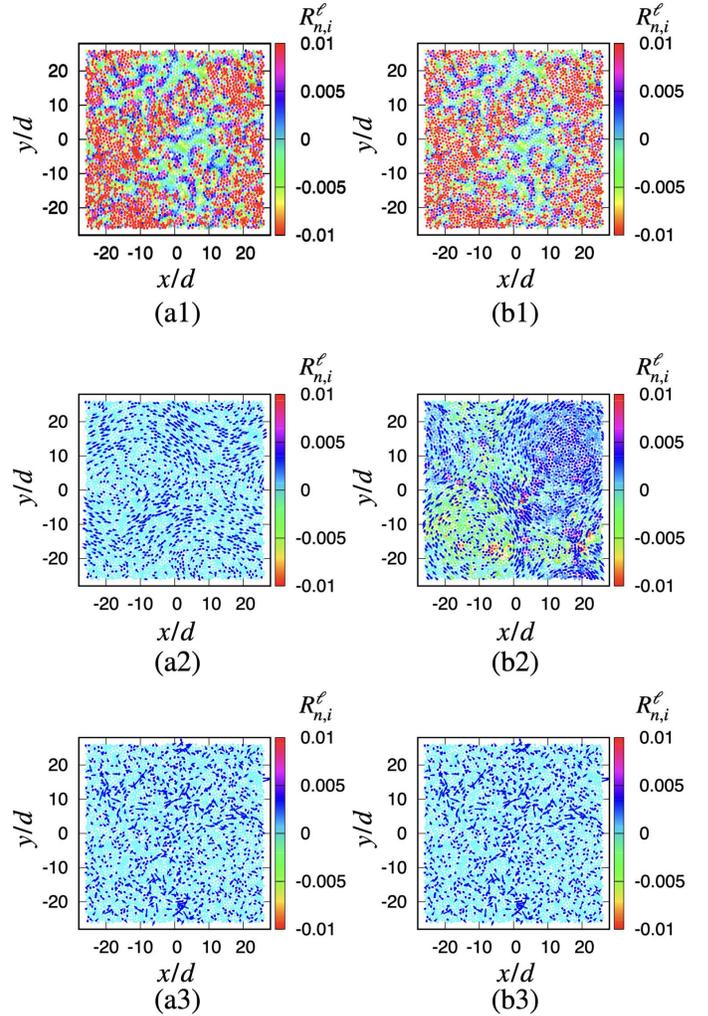}
\caption{
Plots of eigenvectors for $\phi=0.90$ with
(a1) $\omega_{n}t_{0}=1.0\times10^{-4}$,
(a2) $1.3\times10^{-2}$,
(a3) $1.0$,
(b1) $1.0\times10^{-2}$,
(b2) $1.3\times10^{-2}$, and
(b3) $1.0$,
where (a1)-(a3), and (b1)-(b3) are the results for $k_{T}/k_{N}=1.0\times10^{-8}$ and $1.0\times10^{-4}$, respectively.
Here, $(R_{n,i}^x,R_{n,i}^y)^{\textrm{T}}$ and $R_{n,i}^\ell$ are represented by vectors and colors for the $i$-th particle, respectively.
Note that the magnitudes of the vectors are magnified by $50$ times for visualization.
These figures are based on numerical results for $N=4096$.
}
\label{figEiVec}
\end{figure}

In Fig. \ref{figEiVec} we present some typical right eigenvectors $\ket{R_{n}}$,
which was introduced in Eq. \eqref{REieq}, and can be expressed as:
\begin{align}
\ket{R_{n}}=
\left[
\begin{matrix}
\bm{R}_{n,1} \\
\bm{R}_{n,2} \\
\vdots \\
\bm{R}_{n,N}
\end{matrix}
\right]
,
\end{align}
where $\bm{R}_{n,i}:=(R_{n,i}^x,R_{n,i}^y,R_{n,i}^\ell)^{\textrm{T}}$.
Figure \ref{figEiVec} illustrates vectors $(R_{n,i}^x,R_{n,i}^y)^{\textrm{T}}$ and colors to characterize the rotation $R_{n,i}^{\ell}$ of particle $i$ for some characteristic $\omega_n$ with $k_T/k_N=1.0\times 10^{-8}$ (Figs. \ref{figEiVec} (a1)-(a3)) and $k_T/k_N=1.0\times 10^{-4}$ (Figs. \ref{figEiVec} (b1)-(b3)).
Figures \ref{figEiVec} (a1) and (b1) show the eigenvectors at $\omega_{n}t_{0}=1.0\times10^{-2}$ and $\omega_{n}t_{0}=1.0\times10^{-4}$, respectively, which are dominated by the rotational modes.
In Fig. \ref{figEiVec} (a2), we confirm that the eigenvector at $\omega_{n}t_{0}=1.3\times10^{-2}$ is expressed only by translational modes,
whereas the eigenvector at $\omega_{n}t_{0}=1.3\times10^{-2}$ shown in Fig. \ref{figEiVec} (b2) is expressed as a coupling mode of the rotational and translational modes.
In Figs. \ref{figEiVec} (a3) and (b3), we show the eigenvectors at $\omega_{n}t_{0}=1.0$ which are dominated by the translational modes.

\begin{figure}[htbp]
\centering
\includegraphics[width=9cm]{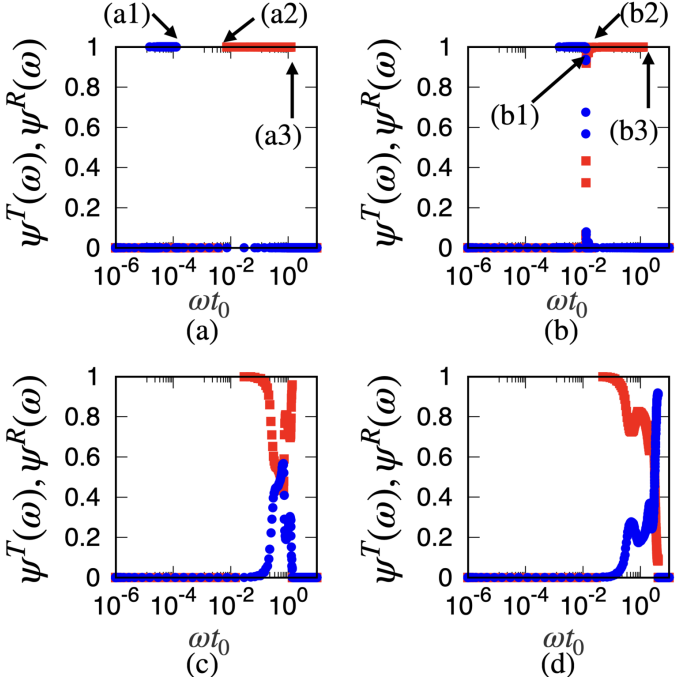}
\caption{
Semi-logarithmic plots of $\psi^{T}$ (red filled squares) and $\psi^{R}$ (blue filled circles) against $\omega t_0$ for $\phi=0.90$
at (a) $k_{T}/k_{N}=1.0\times10^{-8}$,
(b) $1.0\times10^{-4}$,
(c) $1.0$, and
(d) $10.0$.
The eigenvalues of (a1), (a2), (a3), (b1), (b2), and (b3) in these figures correspond to Figs. \ref{figEiVec} (a1), (a2), (a3), (b1), (b2), and (b3), respectively.
These figures are based on numerical results for $N=4096$.
}
\label{figPF}
\end{figure}

To clarify the translational and rotational contributions at each eigenvalue,
we compute the translational and rotational participation fractions \cite{Yunker11,Shiraishi19} defined as
\begin{align}
\psi_{n}^{T}&:=\sum_{i=1}^{N}\left[ |R_{n,i}^{x}|^{2}+|R_{n,i}^{y}|^{2} \right]
\label{PFV},
\\
\psi_{n}^{R}&:=\sum_{i=1}^{N}|R_{n,i}^{\ell}|^{2}
=1-\psi_{n}^{T}
\label{PFR}
,
\end{align}
respectively, where we have investigated the localization of the system with the participation ratio in Appendix \ref{appRatt}.
Note that translational mode is dominant when $\psi_{n}^{T}$ is close to $1$ and rotational mode is dominant when $\psi_{n}^{R}$ is close to $1$.
$\psi^{T}(\omega)$ and $\psi^{R}(\omega)$ are plotted for various $k_{T}/k_{N}$ in Fig. \ref{figPF}, where
\begin{align}
\psi^{T}(\omega):&=\frac{\sum_{n}\nolimits' \langle \psi_{n}^{T}\delta(\omega-\omega_{n}) \rangle}{\sum_{n}\nolimits' \langle\delta(\omega-\omega_{n})\rangle} , \\
\psi^{R}(\omega):&=\frac{\sum_{n}\nolimits' \langle \psi_{n}^{R}\delta(\omega-\omega_{n}) \rangle}{\sum_{n}\nolimits' \langle\delta(\omega-\omega_{n})\rangle}.
\end{align}
Here, $\psi^{T}$ and $\psi^{R}$ are set to zero
if there is no right eigenvalue for $\omega^{(s)}< \omega t_{0}< \omega^{(s+1)}$ with the $s$-th data point $\omega^{(s)}$.
Here, we have used the following steps to determine each data point.
First, we divided the data interval into $50$ parts on a logarithmic scale.
Then we linearly re-divided the data interval from the highest frequency to the 10th highest frequency region.
Finally, we also linearly re-divided the data interval of the log scale corresponding to $0.1\sqrt{k_{T}/k_{N}} <\omega t_{0}<2\sqrt{k_{T}/k_{N}}$.
Note that for the linear re-division of the data, the regions were divided into $500$ or $100$ equally spaced inter-regional intervals for high frequency or $0.1\sqrt{k_{T}/k_{N}} <\omega t_{0}<2\sqrt{k_{T}/k_{N}}$, respectively. 

As shown in Figs. \ref{figPF} (a) and (b),
we find the region of $\psi^R\simeq 1$ for low $\omega$ and $k_T/k_N<1.0\times10^{-4}$.
This region is referred to as Region I.
We also find a region that satisfies $\psi^{T}\simeq1$ for high $\omega$ and $k_T/k_N<1.0$, in which the translational modes are dominant.
This region is referred to as Region II.
Here, three characteristic behaviors depend on $k_{T}/k_{N}$ at $\phi=0.90$.
First, the translational modes are separable from the rotational modes for $k_T/k_N\leq1.0\times10^{-8}$ because we need a small amount of energy to excite the rotational mode in nearly frictionless situations, as shown in Figs. \ref{figPF} (a) and (b).
Second, the translational and rotational contributions are not separated for $1.0\times 10^{-6}\leq k_{T}/k_{N}\leq1.0\times10^{-2}$.
Third, the translational and rotational contributions are indistinguishable for $k_{T}/k_{N}\geq1.0$.

\begin{figure}[htbp]
\centering
\includegraphics[width=9cm]{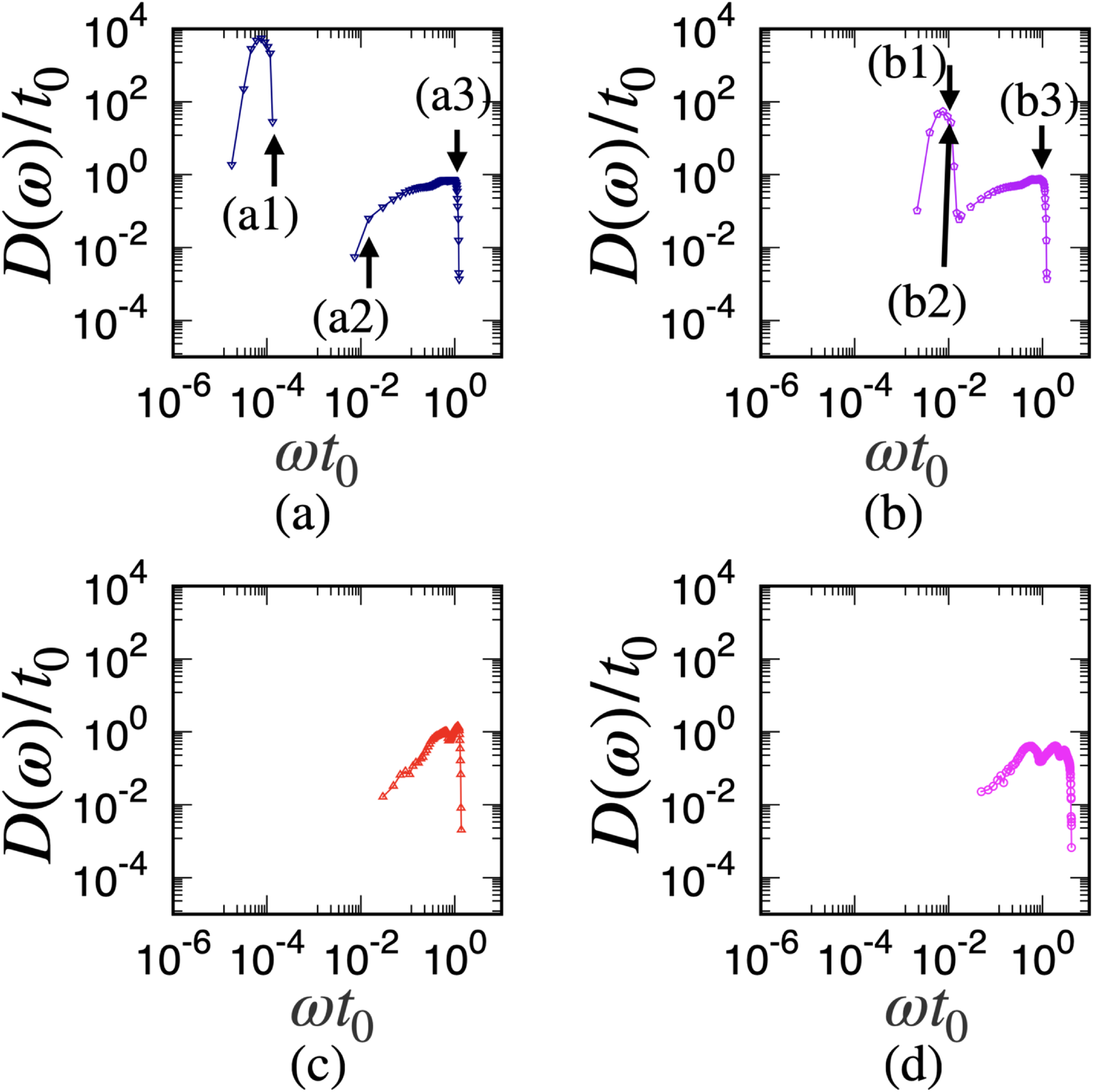}
\caption{
Double logarithmic plots of $D(\omega)$ against $\omega t_0$ for $\phi=0.90$
at (a) $k_{T}/k_{N}=1.0\times10^{-8}$,
(b) $1.0\times10^{-4}$,
(c) $1.0$, and
(d) $10.0$.
The eigenvalues of (a1), (a2), (a3), (b1), (b2) and (b3) in these figures correspond to Figs. \ref{figEiVec} (a1), (a2), (a3), (b1), (b2) and (b3), respectively.
These figures are based on numerical results for $N=4096$.
}
\label{figDOS}
\end{figure}

The DOS obtained from the Jacobian eigenvalues is shown in Fig. \ref{figDOS}.
Based on the results of $\psi^{T}(\omega)$ and $\psi^{R}(\omega)$,
the DOS is also separated into two regions for $k_{T}/k_{N}\leq1.0\times 10^{-8}$.
The rotational band for low $\omega$ shifts to the high $\omega$ region as $k_{T}/k_{N}$ increases (see Figs. \ref{figDOS} (a) and (b)).
In Region II with a high $\omega$ (see Figs. \ref{figDOS} (a) and (b)), the DOS is almost independent of $k_{T}/k_{N}$ in which the translational modes are dominant for $k_{T}/k_{N}\leq1.0\times10^{-2}$.
The distinctions between the two regions for the DOS are visible with a distinct gap between the adjacent regions for $1.0\times10^{-10}\leq k_{T}/k_{N}\leq1.0\times10^{-8}$.
For $1.0\times10^{-6}\leq k_{T}/k_{N}\leq1.0\times10^{-2}$; however, the high $\omega$ region of the DOS in Region I partially overlaps with the low $\omega$ region of Region II.
Furthermore, Regions I and II are completely merged for $k_T/k_N\ge 1.0$ (see Figs. \ref{figDOS} (c) and (d)).
Isolated DOS bands for low $\omega$ have been observed in systems containing anisotropic grains, such as elliptical grains and dimers \cite{Yunker11,Schreck12,Shiraishi19}.
However, to the best of our knowledge, there is no paper pointing out the existence of isolated bands of DOS in systems of frictional grains.

\begin{figure}[htbp]
\centering
\includegraphics[width=8cm]{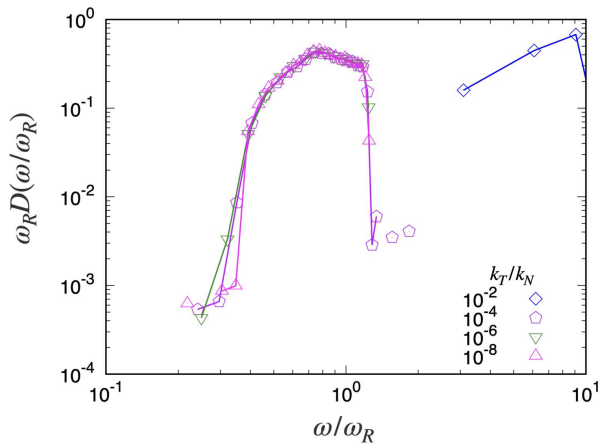}
\caption{
Scaling plots of $\omega_{R} D(\omega)$ versus $\omega/\omega_{R}$ for $\phi=0.90$ and various $k_{T}/k_{N}$ in $0.1<\omega/\omega_{R}<10.0$.
The figure is based on numerical results for $N=4096$.
}
\label{figSclDOS}
\end{figure}

Because we have confirmed the existence of a peak of $D(\omega)$ around $\omega t_{0}\simeq(k_{T}/k_{N})^{1/2}$,
Fig. \ref{figSclDOS} shows the scaling of the DOS in Region I by plotting $\omega_{R} D(\omega/\omega_R)$, where $\omega_{R}:=\sqrt{k_{T}/k_{N}}t_{0}^{-1}$
\footnote{
The reason why $D(\omega)$ is multiplied by $\omega_{R}$ in Fig. \ref{figSclDOS} is as follows.
The integral value of the DOS within Region I $\int_{\textrm{I}}d\omega D(\omega)$ is almost independent of $k_T/k_N$.
Then, LHS can be rewritten as a variable $\int_{\textrm{I}}d\omega D(\omega) = \int_{\textrm{I}}d\hat\omega D^{*}(\hat\omega)$, where $\hat{\omega}:=\omega/\omega_R$, $D^{*}(\hat\omega):=\omega_{R} D(\omega/\omega_{R})$ and $\int_{\textrm{I}}$ represents the integral in Region I.
}.
From Fig. \ref{figSclDOS} we have confirmed that $\omega_{R} D(\omega)$ can be expressed as a universal scaling function of $\omega/\omega_R$ for $0.1<\omega/\omega_R<1$ and $k_T/k_N\le 0.01$.

\begin{figure}[htbp]
\centering
\includegraphics[width=8cm]{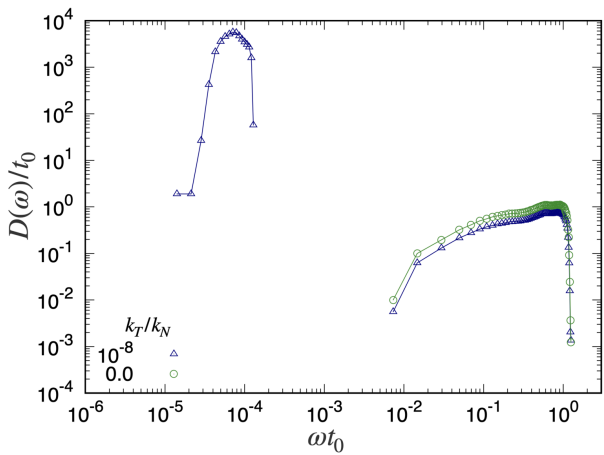}
\caption{
Double logarithmic plots of $D(\omega)/t_{0}$ versus $\omega t_{0}$ with $\phi=0.90$,
where red circles are the DOS for frictional grains, while blue triangles are the DOS for frictionless grains.
The figure is based on numerical results for $N=4096$.
}
\label{DOS0}
\end{figure}

To clarify the behavior of the DOS in the frictionless limit, we compare the DOS for $k_T/k_N=10^{-8}$ with that for frictionless particles by plotting both cases, 
where we adopt the Hessian matrix to calculate the DOS for frictionless systems (see Fig. \ref{DOS0} and Appendix \ref{AppF}).
As expected, there is no singularity of the DOS for the translational mode, while the isolated rotational band in low $\omega$ is absent in frictionless particles, as shown in Fig. \ref{DOS0}.

At the end of this subsection, we examine the usefulness of the effective Hessian $\mH$ introduced in Refs. \cite{Somfai07,Henkes10,Liu21} by comparing $D(\omega)$ with $D_{H}(\omega)$, where $D_{H}(\omega)$ is the DOS obtained from $\mH$ (see Appendix \ref{AppF}).
As shown in Figs. \ref{figDOSH} (a) and (b), $D(\omega)$ and $D_{H}(\omega)$ for various $k_{T}/k_{N}$ are almost identical.
Here, the peak of the DOS near $\omega=0$ is caused by the rotational motion of the grains.
This agreement between the Jacobian and Hessian analyses is natural because the configuration before the application of shear was prepared with frictionless particles, and the tangential displacement $\xi_{T,ij}$ is sufficiently small.

\begin{figure}[htbp]
\centering
\includegraphics[width=8cm]{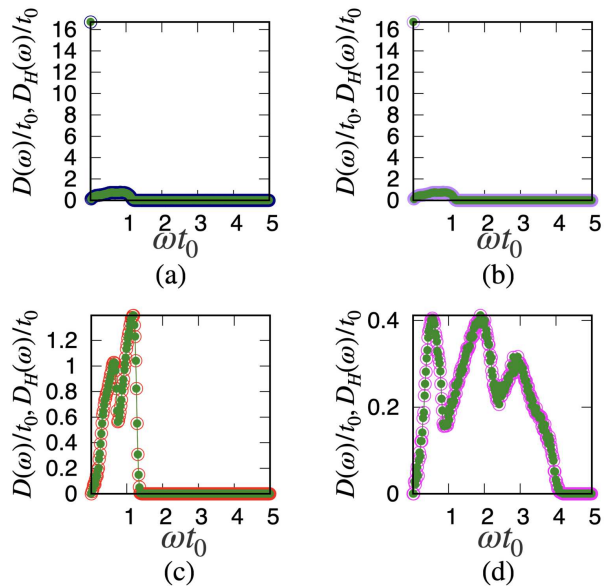}
\caption{
Plots of $D(\omega)$ (filled symbols) and $D_{H}(\omega)$ (open symbols) for $\phi=0.90$ at (a) $k_{T}/k_{N}=1.0\times10^{-8}$, (b) $1.0\times10^{-4}$, (c) $1.0$, and (d) 10.0.
These figures are based on numerical results for $N=4096$.
}
\label{figDOSH}
\end{figure}

\section{Concluding Remarks \label{Discussion}}

We analyzed the eigenmodes of the Jacobian and obtained an expression for the rigidity of amorphous solids of frictional particles under an infinitesimal strain.
We reproduced the rigidity in the linear response regime using the eigenvalues and eigenfunctions of the Jacobian with modifications in the rotational part.

Further, we confirmed that the DOS can be divided into two regions for small $k_{T}/k_{N}$.
In the low-frequency region (Region I), the rotation of the particles plays a dominant role.
These modes are characterized by the frequency $(k_{T}/k_{N})^{1/2}/t_{0}$.
Region I merges into the high-frequency region (Region II) for large $k_T/k_N$, where Region II is dominated by translational modes.

It should be noted that our results are almost independent of system size, as shown in Appendix \ref{SystemSize}. 
Moreover, we have briefly analyzed the density dependence of the DOS in Appendix \ref{DensityDep}, 
where the rotational band shifts to a lower frequency region and the plateau of the translational band become longer as the density approaches the jamming point. 

We have also confirmed that the results of our Jacobian analysis are almost equivalent to those of the Hessian matrix.
This is because our preparation of the initial configuration of grains is made by frictionless grains.
Nevertheless, we expect that the Hessian analysis might be sufficient for stable configurations of grains.


However, the applicability of this theory is limited.
The method used in this study cannot be used for finite strains because it is obvious that the eigenvectors are not orthogonal in the sheared state.
Moreover, there are plastic deformations of the grains under large strains, which were not considered in this study.
Therefore, we cannot predict the correct value of the theoretical rigidity at the stress drop point.
More importantly, the effect of the history dependence of the frictional force is significant even in the linear response regime,
although we have ignored such contacts because of the difficulty in constructing a proper theory.
This issue should be addressed in future studies \cite{Ishima22}.

\section{ACKNOWLEDGMENTS}
The authors thank N. Oyama and T. Kawasaki for fruitful discussions and useful comments.
This work was partially supported by a Grant-in-Aid from the MEXT for Scientific Research (Grant No. JP22K03459, JP21H01006, and JP19K03670) and by the Grant-in-Aid from the Japan Society for Promotion of Science JSPS Research Fellow (Grant No. JP20J20292).

\begin{widetext}

\appendix

\setcounter{equation}{0}

\section{Method of preparing configuration before applying the shear\label{AppIni}}
In this Appendix, we summarize the method of preparing a stable configuration of grains before applying the shear.
For this purpose, as the first step, we perform the relaxation for frictionless particles with the FIRE.
As the second step, the system is relaxed taking into account the static friction between particles.
In the first subsection, we summarize how to prepare the configurations of frictionless particles by the FIRE \cite{Bitzek06}.
In the second subsection, we describe the details of the numerical method including the force with static friction.

\subsection{Method of preparing configuration before applying shear by FIRE}
At first, we place particles at random without any overlaps of particles with the initial fraction $\phi_{\textrm{ini}}=0.6$.
We increase the projected area fraction of the system by the increment of the fraction $\Phi:=\phi^{\textrm{New}}-\phi^{\textrm{Old}}$ up to the target fraction $\phi$,
where $\phi^{\textrm{Old}}$ and $\phi^{\textrm{New}}$ are the projected area fraction of the system before and after each step of the increment, respectively.
After each step of the increment, the system is relaxed by the FIRE \cite{Bitzek06}.

To implement the process of increasing the area fraction, we scale the system as
\begin{align}
L^{\textrm{New}}&=L^{\textrm{Old}}\sqrt{\frac{\phi^{\textrm{Old}}}{\phi^{\textrm{New}}}}, \label{prs1} \\
\bm{r}_{i}^{\textrm{New}}&=\bm{r}_{i}^{\textrm{Old}}\sqrt{\frac{\phi^{\textrm{Old}}}{\phi^{\textrm{New}}}}\label{prs2} ,
\end{align}
where $L^{\textrm{Old}}/L^{\textrm{New}}$ and $\bm{r}^{\textrm{Old}}_{i}/\bm{r}^{\textrm{New}}_{i}$are the linear system size and the position of the $i$-th particle before/after rescaling, respectively.
We adopt $\Phi=10^{-4}$.
When there are overlaps between particles at $\phi^{\textrm{New}}$, the system relaxes to a stable configuration with the aid of the FIRE.

The FIRE is a fast relaxation method of minimizing potentials $U(\bm{r})$ depending on the configuration of the particles $\bm{r}:=(\bm{r}_{1}^{\textrm{T}},\bm{r}_{2}^{\textrm{T}},\cdots,\bm{r}_{N}^{\textrm{T}})^{\textrm{T}}$ with $\bm{r}_{i}:=(r_{i}^{x},r_{i}^{y})^{\textrm{T}}:=(x_{i},y_{i})^{\textrm{T}}$ \cite{Bitzek06}.
Here, we use the Hertzian potential for $U(\bm{r})$ which is defined as
\begin{align}
U(\bm{r}):=\frac{2}{5}k_{N}\sum_{j\neq i}\xi_{N,ij}^{5/2}\Theta(d_{ij}/2-|\bm{r}_{ij}|)
.
\end{align}
Let us introduce $\zeta$-component of the force $F_{F,i}^{\zeta}$ acting on the $i$-th particle as
\begin{align}
F_{F,i}^{\zeta}:&=-\frac{\partial U}{\partial r_{i}^{\zeta}}
=\sum_{j\neq i}k_{N}\xi_{N,ij}^{3/2}n_{ij}^{\zeta}\Theta(d_{ij}/2-|\bm{r}_{ij}|)
,
\end{align}
where $\zeta=x$ or $y$.
Note that $F_{F,i}^{\zeta}$ only consists of the normal repulsive force.
In the FIRE the position $\bm{r}$ and velocity $\bm{v}_{F}:=(\bm{v}_{F,1},\bm{v}_{F,2},\cdots,\bm{v}_{F,N})^{\textrm{T}}$ with $\bm{v}_{F,i}:=(v_{F,i}^{x},v_{F,i}^{y})^{\textrm{T}}$ are updated by the following rules from (i) to (iv) with the variable time increment $\Delta t_{F}$.
(i) The numerical integration via the velocity Verlet method is performed on $\bm{r}$ and $\bm{v}_{F}$:
\begin{align}
r_{i}^{\zeta}&\to r_{i}^{\zeta} +\Delta t_{F} v_{i}^{\zeta} + \Delta t_{F}^{2}\frac{F_{F,i}^{\zeta}(\bm{r})}{2m_{i}}
,
\label{A3}
\\
v_{F,i}^{\zeta}&\to v_{F,i}^{\zeta} + \Delta t_{F}\frac{F_{F,i}^{\zeta}(\tilde{\bm{r}}) + F_{F,i}^{\zeta}(\bm{r}) }{2m_{i}}
,
\end{align}
where $\tilde{\bm{r}}$ is the updated configuration in Eq. \eqref{A3}.
(ii) We calculate $P:=\bm{F}_{F}\cdot\bm{v}_{F}$,
where $\bm{F}_{F}:=(\bm{F}_{F,1}^{\textrm{T}},\bm{F}_{F,2}^{\textrm{T}},\cdots,\bm{F}_{F,N}^{\textrm{T}})^{\textrm{T}}$ with $\bm{F}_{F,i}:=(F_{F,i}^{x},F_{F,i}^{y})^{\textrm{T}}$.
(iii) The velocity $\bm{v}_{F}$ is updated as
\begin{align}
\bm{v}_{F,i}\to\bm{v}_{F,i}+\chi(\hat{\bm{v}}_{F,i}-\hat{\bm{F}}_{F,i})|\bm{v}_{F,i}|,
\end{align}
where $\chi$ is the relaxation parameter and $\hat{\bm{a}}:=\bm{a}/|\bm{a}|$ for an arbitrary vector $\bm{a}$.
(iv) We update $\chi$ and $\Delta t_{F}$ in the FIRE according to the positive or negative value of $P$.
To speed up the relaxation when the motion is along a potential gradient, we increase $\Delta t_{F}$.
Note that this process is performed only when the number of numerical integrations along the potential gradient is larger than a certain number of times $N_{\min}$ to stabilize the numerical calculation.
To implement this update rule, if $P>0$ and the number of numerical integrations of $P>0$ is larger than $N_{\min}$,
$\Delta t_{F}$ and $\chi$ are updated as
\begin{align}
\Delta t_{F}&\to\min(\Delta t_{F} f_{\textrm{inc}},\Delta t_{F,\max}), \\
\chi&\to\chi f_{\chi},
\end{align}
where $\min(a,b)$ is a selecting function of smaller one from $a$ and $b$, the parameter $f_{\textrm{inc}}$ is introduced to speed up the relaxation, and $f_{\chi}$, and $\Delta t_{F,\max}$ are parameters to stabilize numerical calculations.
Here, we adopt $f_{\textrm{inc}}=1.1, f_{\chi}=0.99, N_{\min}=5, \Delta t_{F,\max}=10\Delta t_{F,\textrm{ini}}$, and $\Delta t_{F,\textrm{ini}}=1.0\times10^{-2}t_{0}$ \cite{Bitzek06,Saitoh19}.
Note that $N_{\min}$ is necessary for the stability of the algorithm.
In the case of $P\leq0$, we set
\begin{align}
\bm{v}_{F}&\to\bm{0}, \\
\chi&\to\chi_{\textrm{start}}, \\
\Delta t_{F}&\to\Delta t_{F}f_{\textrm{dec}}
,
\end{align}
where we adopt $f_{\textrm{dec}}=0.5$ and $\chi_{\textrm{start}}=0.1$ \cite{Bitzek06,Saitoh19}.

We repeat the operations (i) through (iv) until $|F_{F,i}^{\zeta}| < F_{\textrm{Th}}$ for arbitrary $i$ and $\zeta$.
Note that we have used the initial values for $\Delta t_{F}=\Delta t_{F,\textrm{ini}}$ and $\chi=\chi_{\textrm{start}}$ at the starting point of the FIRE.
Here, $\bm{r}$ is given and we set $\bm{v}_{F}=\bm{0}$ at the starting point of the FIRE.

\subsection{Numerical method for relaxation of configuration of frictional particles}
After we obtain a stable configuration of frictionless particles at a target fraction in terms of the FIRE,
we consider the effect of static friction in the relaxation process of frictional particles.
The time evolution of the system is given by Eqs. \eqref{tra}-\eqref{fijt} until $\tilde F_{i}^{\alpha}<F_{\textrm{Th}}$ for arbitrary $i$ and $\alpha$.
For the time integration, we adopt the velocity Verlet method with the time increment $\Delta t=1.0\times10^{-2}t_{0}$.

\section{Velocity Verlet method in our system \label{AppVerlet}}
In this Appendix, we first verify the accuracy of the velocity Verlet method.
Next, we summarize the implementation of the velocity Verlet method.
To simplify the notation, we introduce the generalized force $\tilde{\bm{f}}:=(\tilde{\bm{f}}_{1}^{\textrm{T}},\tilde{\bm{f}}_{2}^{\textrm{T}},\cdots,\tilde{\bm{f}}_{N}^{\textrm{T}})^{\textrm{T}}$ with $\tilde{\bm{f}}_{i}:=(\tilde{f}_{i}^{x},\tilde{f}_{i}^{y},\tilde{f}_{i}^{\ell})^{T}:=(\tilde F_{i}^{x}/m_{i}^{x},\tilde F_{i}^{y}/m_{i}^{y},\tilde F_{i}^{\ell}/m_{i}^{\ell})^{\textrm{T}}$ in this Appendix,
where $m_{i}^{\alpha}$ is $m_{i}$ for $\alpha=x,y$ and $4I_{i}/d_{i}^{2}$ for $\alpha=\ell$.
Note that $\tilde{F}_{i}^{\alpha}$ is the generalized force which depends on $\bm{q}$ and $\dot{\bm{q}}$ as in Eqs. \eqref{Fi}-\eqref{fijt}, where $\dot{\bm{a}}:=d\bm{a}/dt$ for an arbitrary vector $\bm{a}$.

\subsection{Accuracy of the velocity Verlet method for the force depending on the velocity}

In this subsection, we check the accuracy of the velocity Verlet method for the force depending on the velocity with the aid of discretization based on the Taylor expansion.
The velocity Verlte method is given by
a set of equations
\begin{align}
q_i^\alpha(t+\Delta t)&=q_i^\alpha(t) + \Delta t\dot{q}_i^\alpha(t) +\frac{1}{2}\Delta t^{2}\tilde{f}_{i}^{\alpha}(t)
\label{(B01)} , \\
\dot{q}_i^\alpha(t+\Delta t)
&=\dot{q}_i^\alpha\left(t\right)
+\Delta t \frac{\tilde{f}_i^\alpha(t) + \tilde{f}_i^\alpha(t+\Delta t)}{2}
\label{(B02)} .
\end{align}
The first equation is called the velocity Velret equation for $q_i^\alpha(t)$ and the second one is the equation for $\dot{q}_i^\alpha(t)$.
It is known that the velocity Verlet algorithm has the accuracy of $O(\Delta t^2)$ in Hamiltonian systems \cite{Swope82},
but the accuracy of this method for dissipative dynamics is little known.
Therefore, we clarify the accuracy of this method in this Appendix.

Here, we show from the Taylor expansion that the velocity Verlet method has a second-order and first-order accuracies of $\Delta t$ for $\bm{q}$ and $\dot{\bm{q}}$, respectively.
Based on the Taylor expansion of $q_{i}^{\alpha}(t+\Delta t)$, we obtain
\begin{align}
q_{i}^{\alpha}(t+\Delta t)
&=q_{i}^{\alpha}(t) + \Delta t\dot{q}_{i}^{\alpha}(t) + \frac{1}{2}\Delta t^{2}\ddot{q}_{i}^{\alpha}(t) + O(\Delta t^{3}) \nonumber \\
&=q_{i}^{\alpha}(t) + \Delta t\dot{q}_{i}^{\alpha}(t) + \frac{1}{2}\Delta t^{2}\tilde{f}_{i}^{\alpha}(t) + O(\Delta t^{3})
\label{qt+Dt}
,
\end{align}
where $\dot{A}:=dA/dt$ for an arbitrary function $A$.
We can obtain the quadratic precision of $\Delta t$ for $q_{i}^{\alpha}(t+\Delta t)$,
because the RHS of Eq. \eqref{qt+Dt} is a function of current time $t$.

On the other hand, based on the Taylor expansion of $\dot{q}_{i}^{\alpha}(t+\Delta t)$, we obtain
\begin{align}
\dot{q}_{i}^{\alpha}(t+\Delta t)
&=\dot{q}_{i}^{\alpha}(t) + \Delta t\ddot{q}_{i}^{\alpha}(t) + \frac{1}{2}\Delta t^{2}\dddot{q}_{i}^{\alpha}(t) + O(\Delta t^{3}) \nonumber \\
&=\dot{q}_{i}^{\alpha}(t) + \Delta t\tilde{f}_{i}^{\alpha}(t) + \frac{1}{2}\Delta t^{2}\dot{\tilde{f}}_{i}^{\alpha}(t) + O(\Delta t^{3})
\label{dotQ}
.
\end{align}
By using $\tilde{f}_{i}^{\alpha}(t+\Delta t)=\tilde{f}_{i}^{\alpha}(t)+\Delta t\dot{\tilde{f}}_{i}^{\alpha}(t)+O(\Delta t^{2})$,
we evaluate $\dot{\tilde{f}}_{i}^{\alpha}(t)$ as
\begin{align}
\dot{\tilde{f}}_{i}^{\alpha}(t)=\frac{\tilde{f}_{i}^{\alpha}(t+\Delta t)-\tilde{f}_{i}^{\alpha}(t)}{\Delta t}+O(\Delta t)
\label{dotF}
.
\end{align}
Substituting Eq. \eqref{dotF} into Eq. \eqref{dotQ}, we obtain \cite{Swope82}
\begin{align}
\dot{q}_{i}^{\alpha}(t+\Delta t)
&=\dot{q}_{i}^{\alpha}(t) + \Delta t\frac{\tilde{f}_{i}^{\alpha}(t+\Delta t) + \tilde{f}_{i}^{\alpha}(t)}{2} + O(\Delta t^{3})
\label{dotQt+Dt}
.
\end{align}
If the force $\tilde{f}_{i}^{\alpha}(t+\Delta t)$ is independent of $\dot{\bm{q}}$,
we can obtain $\tilde{f}_{i}^{\alpha}(\bm{q}(t+\Delta t))$ from $\bm{q}(t+\Delta t)$ with the aid of Eq. \eqref{qt+Dt} \cite{Swope82}.
However, if $\tilde{f}_{i}^{\alpha}(t+\Delta t)$ depends on $\dot{\bm{q}}$,
we have to evaluate $\tilde{f}_{i}^{\alpha}(\bm{q}(t+\Delta t), \dot{\bm{q}}(t+\Delta t) )$,
because $\tilde{f}_{i}^{\alpha}(\bm{q}(t+\Delta t), \dot{\bm{q}}(t+\Delta t) )$ requires LHS of Eq. \eqref{dotQt+Dt}.
Thus, we adopt the following replacements:
\begin{align}
\tilde{f}_{i}^{\alpha}(t+\Delta t)
:&=\tilde{f}_{i}^{\alpha}(\bm{q}(t+\Delta t), \dot{\bm{q}}(t+\Delta t)) 
\to \tilde{f}_{i}^{\alpha}(\bm{q}(t+\Delta t), \dot{\bm{Q}}(t))
,
\label{dQ1}
\\
\tilde{f}_{i}^{\alpha}(t)
:&=\tilde{f}_{i}^{\alpha}(\bm{q}(t), \dot{\bm{q}}(t)) 
\to \tilde{f}_{i}^{\alpha}(\bm{q}(t), \dot{\bm{Q}}(t-\Delta t))
,
\label{dQ2}
\end{align}
where we have introduced
\begin{align}
\dot{Q}_{i}^{\alpha}(t):&=\dot{q}_{i}^{\alpha}(t)+\Delta t\frac{\tilde{f}_{i}^{\alpha}(\bm{q}(t),\dot{\bm{Q}}(t-\Delta t) )}{2},
\\
\dot{Q}_{i}^{\alpha}(t-\Delta t):&=\dot{q}_{i}^{\alpha}(t-\Delta t)+\Delta t\frac{\tilde{f}_{i}^{\alpha}(\bm{q}(t-\Delta t),\dot{\bm{Q}}(t-2\Delta t) )}{2}
.
\end{align}
Here, the difference between $\tilde{f}_{i}^{\alpha}(\bm{q}(t+\Delta t), \dot{\bm{q}}(t+\Delta t))$ and $\tilde{f}_{i}^{\alpha}(\bm{q}(t+\Delta t), \dot{\bm{Q}}(t))$ caused by Eq. \eqref{dQ1} is given by
\begin{align}
\Delta \tilde{f}_{i}^{\alpha}(t+\Delta t)
&:=\tilde{f}_{i}^{\alpha}(\bm{q}(t+\Delta t), \dot{\bm{q}}(t+\Delta t)) - \tilde{f}_{i}^{\alpha}(\bm{q}(t+\Delta t), \dot{\bm{Q}}(t))
\nonumber \\
&= \tilde{f}_{i}^{\alpha}\left( \bm{q}(t+\Delta t), \dot{\bm{q}}(t) + \Delta t\tilde{\bm{f}}(t) + O(\Delta t^{2}) \right)
- \tilde{f}_{i}^{\alpha}\left(\bm{q}(t+\Delta t), \dot{\bm{q}}(t) + \Delta t\frac{\tilde{\bm{f}}( \bm{q}(t), \dot{\bm{q}}(t) )}{2} + O(\Delta t^{2}) \right)
\nonumber \\
&=\Delta t\sum_{j=1}^{N}\sum_{\beta=x,y,\ell}\frac{\tilde{f}_{j}^{\beta}(\bm{q}(t), \dot{\bm{q}}(t))}{2} \frac{\partial \tilde{f}_{i}^{\alpha}(\bm{q}(t), \dot{\bm{q}}(t))}{\partial \dot{q}_{j}^{\beta}} + O(\Delta t^{2})
\label{err1}
.
\end{align}
Similarly, the difference between $\tilde{f}_{i}^{\alpha}(t):=\tilde{f}_{i}^{\alpha}(\bm{q}(t), \dot{\bm{q}}(t))$ and $\tilde{f}_{i}^{\alpha}(\bm{q}(t), \dot{\bm{Q}}(t-\Delta t))$ in Eq. \eqref{dQ2} can be evaluated as
\begin{align}
\Delta \tilde{f}_{i}^{\alpha}(t)
&:=\tilde{f}_{i}^{\alpha}(\bm{q}(t), \dot{\bm{q}}(t)) - \tilde{f}_{i}^{\alpha}(\bm{q}(t), \dot{\bm{Q}}(t-\Delta t))
\nonumber \\
&=\Delta t\sum_{j=1}^{N}\sum_{\beta=x,y,\ell}\frac{\tilde{f}_{j}^{\beta}(\bm{q}(t), \dot{\bm{q}}(t))}{2} \frac{\partial \tilde{f}_{i}^{\alpha}(\bm{q}(t), \dot{\bm{q}}(t))}{\partial \dot{q}_{j}^{\beta}} + O(\Delta t^{2})
\label{err2}
.
\end{align}
Thus, the replacement of Eqs. \eqref{dQ1} and \eqref{dQ2} in Eq. \eqref{dotQt+Dt} with Eqs. \eqref{err1} and \eqref{err2} leads to
\begin{align}
\dot{q}_{i}^{\alpha}(t+\Delta t)
&=\dot{q}_{i}^{\alpha}(t) +\Delta t\frac{ \tilde{f}_{i}^{\alpha}(\bm{q}(t+\Delta t), \dot{\bm{Q}}(t) ) + \tilde{\bm{f}}(\bm{q}(t), \dot{\bm{Q}}(t-\Delta t) ) }{2} \nonumber \\
&\quad- \Delta t^{2}\sum_{j=1}^{N}\sum_{\beta=x,y,\ell}\frac{\tilde{f}_{j}^{\beta}(\bm{q}(t), \dot{\bm{q}}(t))}{2} \frac{\partial \tilde{f}_{i}^{\alpha}(\bm{q}(t), \dot{\bm{q}}(t))}{\partial \dot{q}_{j}^{\beta}}
+ O(\Delta t^{3})
\label{dQt+Dt}
.
\end{align}
Omitting the term including $O(\Delta t^{2})$ in Eq.\eqref{dQt+Dt}, we obtain the following numerical integration methods for $\dot{\bm{q}}$:
\begin{align}
\dot{q}_{i}^{\alpha}(t+\Delta t)
&\to\dot{q}_{i}^{\alpha}(t) +\Delta t\frac{ \tilde{f}_{i}^{\alpha}(\bm{q}(t+\Delta t), \dot{\bm{Q}}(t) ) + \tilde{\bm{f}}(\bm{q}(t), \dot{\bm{Q}}(t-\Delta t) ) }{2}
.
\label{v1}
\end{align}
Note that Eq. \eqref{v1} is the precise expression of the velocity Verlet scheme for $\dot{\bm{q}}$ presented in Eq. \eqref{(B02)}.
From the comparison between Eqs. \eqref{dQt+Dt} and \eqref{v1}, we have confirmed that the velocity Verlet scheme has the first-order precision of $\Delta t$ for $\dot{\bm{q}}$.
We also note that if $\tilde{f}_{i}^{\alpha}$ is independent of $\dot{\bm{q}}$ as in the case of Hamiltonian systems,
the term proportional to $\Delta t^{2}$ is zero and thus, the second-order accuracy of $\Delta t$ for $\dot{q}_{i}^{\alpha}$ is guaranteed.

Let us go back to Eq. \eqref{qt+Dt} with the replacement of Eq. \eqref{dQ2} for $\tilde{f}_{i}^{\alpha}(t)$:
\begin{align}
q_{i}^{\alpha}(t+\Delta t)
&=q_{i}^{\alpha}(t) + \Delta t\dot{q}_{i}^{\alpha}(t) + \frac{1}{2}\Delta t^{2}\tilde{f}_{i}^{\alpha}(t) + O(\Delta t^{3}) \nonumber \\
&=q_{i}^{\alpha}(t) + \Delta t\dot{q}_{i}^{\alpha}(t) + \frac{1}{2}\Delta t^{2}\tilde{f}_{i}^{\alpha}(\bm{q}(t), \dot{\bm{Q}}(t-\Delta t) ) + O(\Delta t^{3})
\label{qt+Dt2}
.
\end{align}
Omitting the term including $O(\Delta t^{3})$ in Eq.\eqref{qt+Dt2}, we obtain the following numerical integration methods for $\bm{q}$:
\begin{align}
q_{i}^{\alpha}(t+\Delta t)
&\to q_{i}^{\alpha}(t) + \Delta t\dot{q}_{i}^{\alpha}(t) + \frac{1}{2}\Delta t^{2}\tilde{f}_{i}^{\alpha}(\bm{q}(t), \dot{\bm{Q}}(t-\Delta t) )
\label{v2}
.
\end{align}
Here, Eq. \eqref{v2} is the precise expression of the velocity Verlet scheme for $\bm{q}$ presented in Eq. \eqref{(B01)}.
From Eqs. \eqref{qt+Dt2} and \eqref{v2} we have confirmed that the velocity Verlet scheme has the second-order precision of $\Delta t$ for $\bm{q}$.

\subsection{Implementation of the velocity Verlet method for the force depending on the velocity}
In this subsection, we explain how to adopt the velocity Verlet method in our system.
In the main text, we adopt the following equations:
\begin{align}
q_{i}^{\alpha}(t+\Delta t) &= q_{i}^{\alpha}(t) + \Delta t\dot{q}_{i}^{\alpha}(t) + \Delta t^{2}\frac{\tilde{f}_{i}^{\alpha}\left( \bm{q}(t), \dot{\bm{Q}}\left(t-\Delta t\right) \right)}{2},
\label{verlet1} \\
\dot{Q}_{i}^{\alpha}\left(t\right) &= \dot{q}_{i}^{\alpha}(t) + \Delta t\frac{\tilde{f}_{i}^{\alpha}\left( \bm{q}(t), \dot{\bm{Q}}\left(t-\Delta t\right) \right)}{2}
\label{verlet2}
.
\end{align}
Here, the updated configuration $q_{i}^{\alpha}(t+\Delta t)$ and modified velocity $\dot{Q}_{i}^{\alpha}\left(t\right)$ are used to obtain the force $\tilde f_{i}^{\alpha}\left( \bm{q}(t+\Delta t), \dot{\bm{Q}}\left(t\right) \right)$.
Then, we update the velocity as follows
\begin{align}
\dot{q}_{i}^{\alpha}(t+\Delta t) &= \dot{Q}_{i}^{\alpha}\left(t\right) + \Delta t\frac{\tilde{f}_{i}^{\alpha}\left( \bm{q}(t+\Delta t), \dot{\bm{Q}}\left(t\right) \right)}{2}
\label{verlet3}
.
\end{align}

\section{Jacobian Properties \label{AppC}}
In this Appendix, we summarize the properties of the Jacobian introduced in Eq. \eqref{Jacobian}.
\subsection{Jacobian block elements}
Let us write $3\times3$ sub-matrix $\mJ_{ij}$, which is $(ij)$ block element of the Jacobian obtained from Eq. \eqref{Jacobian}:
\begin{align}
[\mJ_{ij}]^{\alpha\beta}
&:=-\frac{\partial\tilde{F}_{i}^{\alpha}}{\partial q_{j}^{\beta}} \nonumber \\
&=
\left[
\begin{matrix}
-\partial_{q_{j}^{x}}F_{i}^{x}    &  -\partial_{q_{j}^{y}}F_{i}^{x}  & -\partial_{q_{j}^{\ell}} F_{i}^{x} \\
-\partial_{q_{j}^{x}}F_{i}^{y}    &  -\partial_{q_{j}^{y}}F_{i}^{y}  & -\partial_{q_{j}^{\ell}} F_{i}^{y} \\
-\partial_{q_{j}^{x}}\tilde T_{i} & -\partial_{q_{j}^{y}}\tilde T_{i} & -\partial_{q_{j}^{\ell}}\tilde T_{i}
\end{matrix}
\right]
\nonumber \\
&=
\left[
\begin{matrix}
-\sum_{k=1;k\neq j}^{N}\partial_{q_{j}^{x}}f_{ik}^{x} & -\sum_{k=1;k\neq j}^{N}\partial_{q_{j}^{y}}f_{ik}^{x} & -\sum_{k=1;k\neq j}^{N}\partial_{q_{j}^{\ell}}f_{ik}^{x}\\
-\sum_{k=1;k\neq j}^{N}\partial_{q_{j}^{x}}f_{ik}^{y} & -\sum_{k=1;k\neq j}^{N}\partial_{q_{j}^{y}}f_{ik}^{y} & -\sum_{k=1;k\neq j}^{N}\partial_{q_{j}^{\ell}}f_{ik}^{y}\\
-\sum_{k=1;k\neq j}^{N}\partial_{q_{j}^{x}}\tilde T_{ik} & -\sum_{k=1;k\neq j}^{N}\partial_{q_{j}^{y}}\tilde T_{ik} & -\sum_{k=1;k\neq j}^{N}\partial_{q_{j}^{\ell}}\tilde T_{ik}
\end{matrix}
\right]
\nonumber \\
&=
\begin{cases}
\left[
\begin{matrix}
- \partial_{q_{j}^{x}}f_{ij}^{x} & - \partial_{q_{j}^{y}}f_{ij}^{x} & - \partial_{q_{j}^{\ell}}f_{ij}^{x}\\
- \partial_{q_{j}^{x}}f_{ij}^{y} & - \partial_{q_{j}^{y}}f_{ij}^{y} & - \partial_{q_{j}^{\ell}}f_{ij}^{y}\\
- \partial_{q_{j}^{x}}\tilde T_{ij} & - \partial_{q_{j}^{y}}\tilde T_{ij} & - \partial_{q_{j}^{\ell}}\tilde T_{ij}
\end{matrix}
\right]
(i\neq j)\\
\left[
\begin{matrix}
-\sum_{k=1;k\neq i}^{N}\partial_{q_{i}^{x}}f_{ik}^{x} & -\sum_{k=1;k\neq i}^{N}\partial_{q_{i}^{y}}f_{ik}^{x} & -\sum_{k=1;k\neq i}^{N}\partial_{q_{i}^{\ell}}f_{ik}^{x}\\
-\sum_{k=1;k\neq i}^{N}\partial_{q_{i}^{x}}f_{ik}^{y} & -\sum_{k=1;k\neq i}^{N}\partial_{q_{i}^{y}}f_{ik}^{y} & -\sum_{k=1;k\neq i}^{N}\partial_{q_{i}^{\ell}}f_{ik}^{y}\\
-\sum_{k=1;k\neq i}^{N}\partial_{q_{i}^{x}}\tilde T_{ik} & -\sum_{k=1;k\neq i}^{N}\partial_{q_{i}^{y}}\tilde T_{ik} & -\sum_{k=1;k\neq i}^{N}\partial_{q_{i}^{\ell}}\tilde T_{ik}
\end{matrix}
\right]
(i=j)
\end{cases}
\label{calJac}
,
\end{align}
where the superscripts $\alpha$ and $\beta$ correspond to $x,y,\ell$-components, and $i$ and $j$ are the particle numbers (see Appendix \ref{JacobianHertz} for each component of $\mJ$).
Here, $f_{ij}^{\zeta},\tilde T_{ij}$ are $\zeta$-component of $\bm{f}_{ij}$ and scaled torque that the $i$-th particle receives from the $j$-th particle, respectively.
The sub-matrix  for $i=j$ is given by
\begin{align}
[\mJ_{ii}]^{\alpha\beta}
=
\left[
\begin{matrix}
\sum_{k=1;k\neq i}^{N}\partial_{q_{k}^{x}}f_{ik}^{x} & \sum_{k=1;k\neq i}^{N}\partial_{q_{k}^{y}}f_{ik}^{x} & -\sum_{k=1;k\neq i}^{N}\partial_{q_{k}^{\ell}}f_{ik}^{x}\\
\sum_{k=1;k\neq i}^{N}\partial_{q_{k}^{x}}f_{ik}^{y} & \sum_{k=1;k\neq i}^{N}\partial_{q_{k}^{y}}f_{ik}^{y} & -\sum_{k=1;k\neq i}^{N}\partial_{q_{k}^{\ell}}f_{ik}^{y}\\
\sum_{k=1;k\neq i}^{N}\partial_{q_{k}^{x}}\tilde T_{ik} & \sum_{k=1;k\neq i}^{N}\partial_{q_{k}^{y}}\tilde T_{ik} & -\sum_{k=1;k\neq i}^{N}\partial_{q_{k}^{\ell}}\tilde T_{ik}
\end{matrix}
\right]
\label{calJac2}
,
\end{align}
where we have used $\partial_{q_{i}^{\kappa}}f_{ik}^{\zeta} = -\partial_{q_{k}^{\kappa}}f_{ik}^{\zeta}, \partial_{q_{i}^{\kappa}}\tilde T_{ik} = -\partial_{q_{i}^{\kappa}}\tilde T_{ik}, \partial_{q_{i}^{\ell}}f_{ik}^{\zeta} = \partial_{q_{k}^{\ell}}f_{ik}^{\zeta}$, and $ \partial_{q_{i}^{\ell}}\tilde T_{ik} = \partial_{q_{i}^{\ell}}\tilde T_{ik}$.
Here, the superscripts $\zeta$ and $\kappa$ correspond to $x,y$ components.

From Eqs. \eqref{calJac} and \eqref{calJac2} $\mJ_{ij}^{\zeta\beta}$ satisfies
\begin{align}
\mJ_{ii}^{\zeta\beta}&=-\sum_{j\neq i}\mJ_{ij}^{\zeta\beta} \label{eq10}
\end{align}
Thus, introducing $J_{nm}\ (n,m=1,2,\cdots, 3N)$ which is a rewriting of $\mJ_{ij}^{\alpha\beta}$ in Eq. \eqref{defJ} by the index from $i$ and $\alpha$ to $n$, we obtain
\begin{align}
\sum_{n=1,4,\cdots,3N-2} J_{nm}&=0, \label{Jx0}\\
\sum_{n=2,5,\cdots,3N-1} J_{nm}&=0.
\end{align}
where $\sum_{n=1,4,\cdots,3N-2}$ and $\sum_{n=2,5,\cdots,3N-1}$ express the summations of modulus $1$ and modulus $2$ with the intervals $3$, respectively.
Here, we write $3N$-dimensional vector translating in the $x$ direction $\bm{e}_{x}$ as
\begin{align}
\bm{e}_{x}&=
\left[
\begin{matrix}
\bm{e}_{x,1}\\
\bm{e}_{x,2}\\
\vdots\\
\bm{e}_{x,N}
\end{matrix}
\right].
\end{align}
where $\bm{e}_{x,i}:=(1,0,0)^{\textrm{T}}$ for $i=1,2,\cdots, N$.
Here, the $n$-th component of the action of $\mJ$ on $\bm{e}_{x}$ satisfies
\begin{align}
\{\mJ\bm{e}_{x}\}_{n}&=\sum_{m}J_{nm}e_{x,m} \nonumber \\
&=\sum_{m=1,4,\cdots,3N-2}J_{nm}\nonumber \\
&=0,
\end{align}
where we have used Eq. \eqref{Jx0} for the last equality.
Thus, we obtain $\mJ\bm{e}_{x}=\bm{0}$, where $\bm{0}$ is zero vector.
Similarly, using
\begin{align}
\bm{e}_{y}&=
\left[
\begin{matrix}
\bm{e}_{y,1}\\
\bm{e}_{y,2}\\
\vdots\\
\bm{e}_{y,N}
\end{matrix}
\right]
,
\end{align}
with $\bm{e}_{y,i}:=(0,1,0)^{\textrm{T}}$ we also obtain $\mJ\bm{e}_{y}=\bm{0}$.
Therefore, $\bm{e}_{x}$ and $\bm{e}_{y}$ are the zero modes for $\mJ$.

\section{Explicit Jacobian expressions \label{AppD}}
In this Appendix, we present the explicit expressions of the Jacobian based on Eqs. \eqref{fijSum}-\eqref{fijt}.
Then, we clarify the difference between the present results and the case where the tangential force is approximated by the conservative force used in the previous studies \cite{Somfai07,Henkes10}.

\subsection{Calculation of Jacobian }
Let us consider only the normal and tangential elastic contact forces
\begin{align}
\bm{f}_{N,ij}&=k_{N}\xi_{N,ij}^{3/2}\bm{n}_{ij}, \label{Hertzian2}\\
\bm{f}_{T,ij}&=k_{T}\xi_{N,ij}^{1/2}\bm{\xi}_{T,ij} \label{Hertzian3}
,
\end{align}
where the integration of $d\bm{\xi}_{T,ij}$
\begin{align}
\bm{\xi}_{T,ij}:=\int_{C_{ij}}d\bm{\xi}_{T,ij} \label{xiTint}
\end{align}
is performed during the contact between $i$ and $j$ particles.
Since Eq. \eqref{xiTint} does not contain the second term on RHS of Eq. \eqref{xiT}, $\bm{\xi}_{T,ij}$ may not be perpendicular to $\bm{\xi}_{N,ij}$.
Neverthelss, we adopt Eq. \eqref{xiTint} for simplicity. 
Here, $d{\bm{\xi}}_{T,ij}$ is defined as
\begin{align}
d{\bm{\xi}}_{T,ij}= d{\bm{r}}_{ij}-(d{\bm{r}}_{ij}\cdot\bm{n}_{ij})\bm{n}_{ij}-d{\bm{\ell}}_{ij}\times\bm{n}_{ij}
\label{velotangential}
,
\end{align}
where $\bm{\ell}_{ij}$ is defined as
\begin{align}
\bm{\ell}_{ij}
:=
\left[
\begin{matrix}
0\\
0\\
\ell_{i}+\ell_{j}
\end{matrix}
\right].
\end{align}
Each component of Eq. \eqref{velotangential} is written as
\begin{align}
d{\xi}_{T,ij}^{x}
&= d{{r}}_{ij}^{x}-(d{\bm{r}}_{ij}\cdot\bm{n}_{ij}){n}_{ij}^{x}+d{{\ell}}_{ij}{n}_{ij}^{y}
, \\
d{\xi}_{T,ij}^{y}
&= d{{r}}_{ij}^{y}-(d{\bm{r}}_{ij}\cdot\bm{n}_{ij}){n}_{ij}^{y}-d{{\ell}}_{ij}{n}_{ij}^{x}
.
\end{align}

The derivative of the normal force is given by
\begin{align}
\partial_{r_{i}^{\zeta}}f_{N,ij}^{\kappa}
&=k_{N}\left[
\delta_{\zeta\kappa}\frac{\xi_{N,ij}^{3/2}}{r_{ij}} - \left( \frac{3}{2}+\frac{\xi_{N,ij}}{r_{ij}} \right)\xi_{N,ij}^{1/2}n_{ij}^{\zeta}n_{ij}^{\kappa}
\right],
\label{N1st}
\\
\partial_{\ell_{i}}f_{N,ij}^{\kappa}&=0,
\end{align}
where Kronecker's delta $\delta_{\zeta\kappa}$ satisfies $\delta_{\zeta\kappa}=1$ for $\zeta=\kappa$ and $\delta_{\zeta\kappa}=0$ otherwise.
We have used
\begin{align}
\frac{\partial n_{ij}^{\zeta}}{\partial r_{i}^{\kappa}}&=\frac{1}{r_{ij}}\left( \delta_{\zeta\kappa}-n_{ij}^{\zeta}n_{ij}^{\kappa}\right),\\
\frac{\partial r_{ij}}{\partial r_{i}^{\zeta}}&=n_{ij}^{\zeta}
\end{align}
to obtain Eq. \eqref{N1st}.

The derivative of the tangential force is written as
\begin{align}
\partial_{r_{i}^{\zeta}}f_{T,ij}^{\kappa}
&= \frac{1}{2}k_{T}\xi_{N,ij}^{-1/2}n_{ij}^{\zeta}\xi_{T,ij}^{\kappa}
- k_{T}\xi_{N,ij}^{1/2}\left( \delta_{\zeta\kappa}-n_{ij}^{\zeta}n_{ij}^{\kappa} \right)
,
\label{T1st}
\\
\partial_{\ell_{i}}f_{T,ij}^{\kappa}
&=-\varepsilon_{\kappa}k_{T}\xi_{N,ij}^{1/2}n_{ij}^{\nu_{\kappa}}
\label{ellFt},
\end{align}
where $\varepsilon_{\zeta}$ and $\nu_{\zeta}$ are, respectively, defined as
\begin{align}
\varepsilon_{\zeta}:&=
\left\{
\begin{matrix}
1 \quad (\zeta=x)\\
-1 \quad (\zeta=y),
\end{matrix}
\right.
\\
\nu_{\zeta}:&=
\left\{
\begin{matrix}
y \quad (\zeta=x)\\
x \quad (\zeta=y).
\end{matrix}
\right.
\end{align}

Here, $\partial_{r_{i}^{\zeta}} {\xi}_{T,ij}^{\kappa}$ and $\partial_{\ell_{i}} {\xi}_{T,ij}^{\kappa} $ in Eqs. \eqref{T1st} and \eqref{ellFt} satisfy
\begin{align}
\frac{\partial  {\xi}_{T,ij}^{\kappa}  }{\partial r_{i}^{\zeta}}
&=\delta_{\zeta\kappa}-n_{ij}^{\zeta}n_{ij}^{\kappa},\label{eq45}
\\
\frac{\partial  {\xi}_{T,ij}^{\kappa}  }{\partial \ell_{i}}
&=  \varepsilon_{\kappa} n_{ij}^{\nu_{\kappa}}
.
\label{eq46}
\end{align}
The derivation of Eqs. \eqref {eq45} and \eqref {eq46} are as follows \cite{Chattoraj19}.
From Eq. \eqref{velotangential} $d {\xi}_{T,ij}^{\zeta}$ can be written as
\begin{align}
d {\xi}_{T,ij}^{\zeta}  = dr_{ij}^{\zeta}-(d\bm{r}_{ij}\cdot\bm{n}_{ij})n_{ij}^{\zeta} + (-1)^{\zeta}(d\ell_{i}+d\ell_{j})n_{ij}^{\nu_{\zeta}}
\label{eq23}
.
\end{align}
Then, $d\xi_{T,ij}^{x}$ satisfies
\begin{align}
d{\xi}_{T,ij}^{x} &= dr_{ij}^{x}-\sum_{\kappa=x,y}dr_{ij}^{\kappa}n_{ij}^{\kappa}n_{ij}^{x} + n_{ij}^{y}(d\ell_{i}+d\ell_{j})
\nonumber \\
&=(1-(n_{ij}^{x})^{2})dr_{ij}^{x} - n_{ij}^{x}n_{ij}^{y}dr_{ij}^{y} + n_{ij}^{y}(d\ell_{i}+d\ell_{j})
\nonumber \\
&= (n_{ij}^{y})^{2}dr_{ij}^{x} - n_{ij}^{x}n_{ij}^{y}dr_{ij}^{y} + n_{ij}^{y}(d\ell_{i}+d\ell_{j})
\nonumber \\
&= (n_{ij}^{y})^{2}(dx_{i}-dx_{j}) - n_{ij}^{x}n_{ij}^{y}(dy_{i}-dy_{j}) + n_{ij}^{y}(d\ell_{i}+d\ell_{j})
\label{dxitx}.
\end{align}
Similarly, $d\xi_{T,ij}^{y}$ also satisfies
\begin{align}
d\xi_{T,ij}^{y}
&= -n_{ij}^{x}n_{ij}^{y}(dx_{i}-dx_{j}) + (n_{ij}^{y})^{2}(dy_{i}-dy_{j}) - n_{ij}^{x}(d\ell_{i}+d\ell_{j})
\label{dxity}
.
\end{align}
Here, $d{\xi}_{T,ij}^{\zeta}$is the function of $x_{i}, y_{i}, \ell_{i}, x_j, y_j,$ and $\ell_{j}$.
We obtain the differential form of $d{\xi}_{T,ij}^{\zeta}$:
\begin{align}
d {\xi}_{T,ij}^{\zeta}
&=
  \left(\frac{\partial {\xi}_{T,ij}^{\zeta}}{\partial x_{i}}\right)_{(y_{i},\ell_{i},x_{j},y_{j},\ell_{j})}dx_{i}
+\left(\frac{\partial {\xi}_{T,ij}^{\zeta}}{\partial x_{j}}\right)_{(x_{i},y_{i},\ell_{i},y_{j},\ell_{j})}dx_{j}
\nonumber \\
&\quad
+\left(\frac{\partial {\xi}_{T,ij}^{\zeta}}{\partial y_{i}}\right)_{(x_{i},\ell_{i},x_{j},y_{j},\ell_{j})}dy_{i}
+\left(\frac{\partial {\xi}_{T,ij}^{\zeta}}{\partial y_{j}}\right)_{(x_{i},y_{i},\ell_{i},x_{j},\ell_{j})}dy_{j}
\nonumber \\
&\quad
+\left(\frac{\partial {\xi}_{T,ij}^{\zeta}}{\partial \ell_{i}}\right)_{(x_{i},y_{i},x_{j},y_{j},\ell_{j})}d\ell_{i}
+\left(\frac{\partial {\xi}_{T,ij}^{\zeta}}{\partial \ell_{j}}\right)_{(x_{i},y_{i},\ell_{i},x_{j},y_{j})}d\ell_{j}
\label{dxit}
.
\end{align}
Then, we obtain Eqs. \eqref{eq45}, \eqref{eq46}, by comparing Eqs. \eqref {dxitx} and \eqref {dxity} with Eq. \eqref {dxit}.

Since the scaled torque $\tilde T_{ij}$ satisfies
\begin{align}
\tilde{T}_{ij}:=\frac{2T_{ij}}{d_{i}}=-n_{ij}^{x}f_{T,ij}^{y} + n_{ij}^{y} f_{T,ij}^{x}
,
\end{align}
we obtain
\begin{align}
\partial_{r_{i}^{\zeta}}\tilde{T}_{ij}
&=-\left(\partial_{r_{i}^{\zeta}}n_{ij}^{x}\right)f_{T,ij}^{y} -n_{ij}^{x}\partial_{r_{i}^{\zeta}}f_{T,ij}^{y} + \left(\partial_{r_{i}^{\zeta}}n_{ij}^{y}\right) f_{T,ij}^{x} + n_{ij}^{y} \partial_{r_{i}^{\zeta}}f_{T,ij}^{x}
\nonumber \\
&=-\left( \frac{\delta_{\zeta x}}{r_{ij}}-\frac{n_{ij}^{\zeta}n_{ij}^{x}}{r_{ij}}\right)f_{T,ij}^{y}
-n_{ij}^{x}
\left[
\frac{1}{2}k_{T}\xi_{N,ij}^{-1/2}\xi_{T,ij}n_{ij}^{\zeta}t_{ij}^{y}
- k_{T}\xi_{N,ij}^{1/2}\left( \delta_{\zeta y}-n_{ij}^{\zeta}n_{ij}^{y} \right)
\right]
\nonumber \\
\nonumber \\
&\quad+\left( \frac{\delta_{\zeta y}}{r_{ij}}-\frac{n_{ij}^{\zeta}n_{ij}^{y}}{r_{ij}}\right)f_{T,ij}^{x}
+n_{ij}^{y}
\left[
\frac{1}{2}k_{T}\xi_{N,ij}^{-1/2}\xi_{T,ij}n_{ij}^{\zeta}t_{ij}^{x}
- k_{T}\xi_{N,ij}^{1/2}\left( \delta_{\zeta x}-n_{ij}^{\zeta}n_{ij}^{x} \right)
\right]
\nonumber \\
&=
-n_{ij}^{x}
\left[
\frac{1}{2}k_{T}\xi_{N,ij}^{-1/2}\xi_{T,ij}n_{ij}^{\zeta}t_{ij}^{y}
- k_{T}\xi_{N,ij}^{1/2}\left( \delta_{\zeta y}-n_{ij}^{\zeta}n_{ij}^{y} \right)
\right]
\nonumber \\
&\quad
+n_{ij}^{y}
\left[
\frac{1}{2}k_{T}\xi_{N,ij}^{-1/2}\xi_{T,ij}n_{ij}^{\zeta}t_{ij}^{x}
- k_{T}\xi_{N,ij}^{1/2}\left( \delta_{\zeta x}-n_{ij}^{\zeta}n_{ij}^{x} \right)
\right]
\quad
\nonumber \\
&=
-n_{ij}^{x}
\left[
\frac{1}{2}k_{T}\xi_{N,ij}^{-1/2}\xi_{T,ij}n_{ij}^{\zeta}t_{ij}^{y}
- k_{T}\xi_{N,ij}^{1/2}\delta_{\zeta y}
\right]
+n_{ij}^{y}
\left[
\frac{1}{2}k_{T}\xi_{N,ij}^{-1/2}\xi_{T,ij}n_{ij}^{\zeta}t_{ij}^{x}
- k_{T}\xi_{N,ij}^{1/2}\delta_{\zeta x}
\right]
\label{xyT}
,
\\
\partial_{\ell_{i}}\tilde{T}_{ij}
&=-n_{ij}^{x}\partial_{\ell_{i}}f_{T,ij}^{y} + n_{ij}^{y}\partial_{\ell_{i}}f_{T,ij}^{x}
\nonumber \\
&=-n_{ij}^{x} k_{T}\xi_{N,ij}^{1/2}n_{ij}^{x}
-n_{ij}^{y} k_{T}\xi_{N,ij}^{1/2}n_{ij}^{y}
\nonumber \\
&=-k_{T}\xi_{N,ij}^{1/2}
\label{last}
,
\end{align}
where we have used $\sum_{\zeta}f_{T,ij}^\zeta n_{ij}^\zeta=0$.

The terms proportional to $\xi_{T,ij}$ in the Jacobian include the history-dependent tangential displacements which are ignored in the effective potential (see Appendix \ref{AppF}) \cite{Somfai07,Henkes10,Liu21}.
The reason we use the Jacobian is to include the history-dependent tangential displacements in the dynamical matrix.

\section{Effects of rattlers \label{appRatt}}
In this Appendix, we investigate the effects of rattlers.
In the first subsection, we investigate the effects of rattlers for the DOS.
In the second subsection, we clarify the contributions of rattlers by using the participation ratio.

\subsection{Effects of rattlers on the DOS\label{appRattDOS} }

In this subsection, we investigate the role of rattlers.
We call particle $i$ a rattler, if its coordination number $Z_i$ is $Z_{i}\leq Z_{\textrm{Th}}$.
Since the coordination number of isostatic state is three, $Z_{\textrm{Th}}$ can be $1$ or $2$ for frictional grains.
The rattlers are determined by the following method.
Given a particle configuration, we measure the coordination number $Z_{i}^{(n=1)}$ of each particle.
Then, we regard $N_{1}$ particles satisfying $Z_{i}^{(n=1)}\leq Z_{\textrm{Th}}$ as rattlers at the first trial.
We measure the coordination number $Z_{i}^{(n=2)}$ after we remove the rattler particles.
In the second trial, we regard particles satisfying $Z_{i}^{(n=2)}\leq Z_{\textrm{Th}}$ as new rattlers.
We repeat these processes until the number of rattlers is converged.
As shown in Fig. \ref{figDOSratt}, at which we adopt $Z_{\textrm{Th}}=2$,
low-frequency modes in Region I and intermediate modes between Regions I and II are contributions from rattlers.
Thus, we conclude that the contributions of rattlers on the DOS are not important. 

\begin{figure}[htbp]
\centering
\includegraphics[width=16cm]{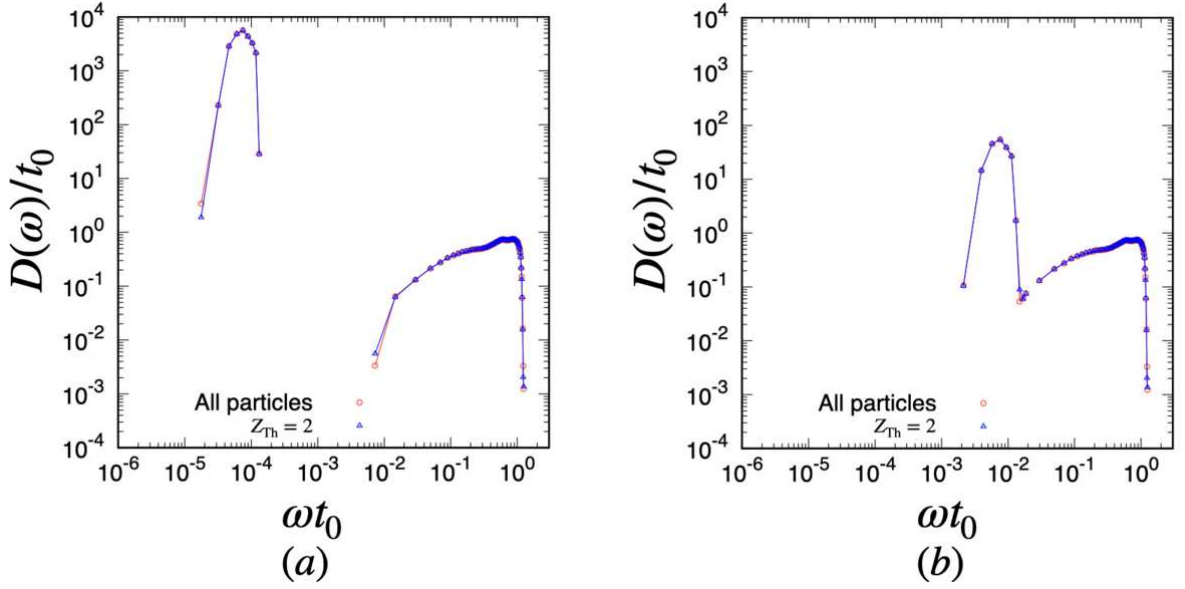}
\caption{
Double logarithmic plots of $D(\omega)$ with (red circles) and without (blue triangles by using $Z_{\textrm{Th}}=2$)  rattlers against $\omega t_0$ for $\phi=0.90$  at (a) $k_{T}/k_{N}=1.0\times10^{-8}$ and (b) $k_{T}/k_{N}=1.0\times10^{-4}$.
These figures are based on numerical results for $N=4096$.
}
\label{figDOSratt}
\end{figure}

\subsection{ Participation ratio}

In this subsection, to clarify whether the mode at $\omega_n$ is localized or spread to the whole system,
we introduce a participation ratio $p_{n}$ \cite{Yunker11,Shiraishi19} 
\begin{align}
p_{n}:=\frac{\left( \sum_{i=1}^{N}|\bm{R}_{n,i}|^{2} \right)^{2}}{N\sum_{i=1}^{N}|\bm{R}_{n,i}|^{4}}
.
\label{PR}
\end{align}
We plot $p(\omega)$ defined as
\begin{align}
p(\omega):=\frac{\sum_{n}\nolimits' \langle p_{n}\delta(\omega-\omega_{n}) \rangle}{\sum_{n}\nolimits' \langle\delta(\omega-\omega_{n})\rangle}
\end{align}
against $\omega t_{0}$ for $\phi=0.90$ in Fig. \ref{figPR}.
Note that $p(\omega)$ are set to be zero if there is no right eigenvalue in the region $(\omega^{(s)}< \omega< \omega^{(s+1)})$.

\begin{figure}[htbp]
\centering
\includegraphics[width=16cm]{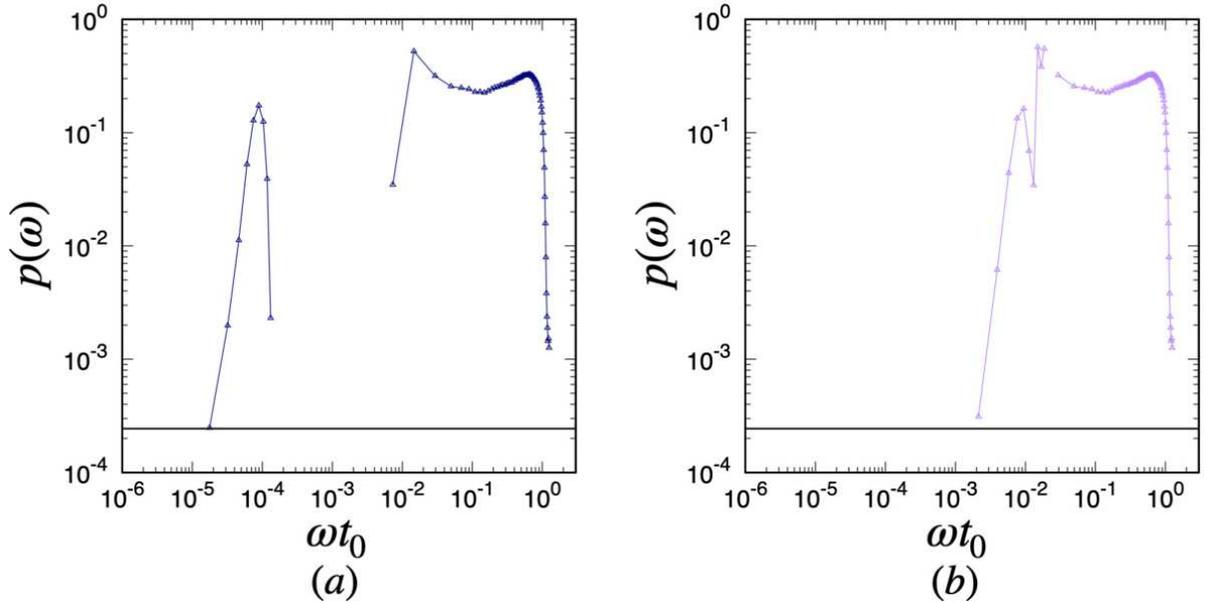}
\caption{
Double logarithmic plots of $p(\omega)$ for $\phi=0.90$ against $\omega t_0$ at (a) $k_{T}/k_{N}=1.0\times10^{-8}$ and (b) $k_{T}/k_{N}=1.0\times10^{-4}$, where the guide line is $1/N$.
These figures are based on numerical results for $N=4096$.
}
\label{figPR}
\end{figure}

Figure \ref{figPR} shows that the modes at $\omega t_{0}\approx10^{-5}$ in Fig. \ref{figPR} (a) and at $\omega t_{0}\approx10^{-3}$ in Fig. \ref{figPR} (b) are nearly equal to $p\approx1/N$.
Recalling that those modes consist of the rattler, we conclude that the contribution of the rattler is localized.
In the middle range of  $\omega$ in Fig. \ref{figPR}, there is an isolated band which shifts to the large $\omega$ as $k_T/k_N$ increases with keeping its shape which can be seen in the main text.

\section{Explicit results of $\mJ_{N}$ and $\mJ_{T}$  \label{JacobianHertz}}

In this Appendix, we have written down the explicit results of $\mJ_{N}$ and $\mJ_{T}$. 
From the results for the derivative of $\tilde{F}_{i}^{\alpha}$ in Appendix D , the non-diagonal block elements $\mJ_{N,ij}^{\alpha\beta}, \mJ_{T,ij}^{\alpha\beta} (i\neq j)$ are given by
\begin{align}
\mJ_{N,ij}^{xx}
&=k_{N}\frac{\xi_{ij,N}^{3/2}}{r_{ij}} - k_{N}\left[ \frac{3}{2}+\frac{\xi_{ij,N}}{r_{ij}} \right]\xi_{ij,N}^{1/2}(n_{ij}^{x})^{2}
\label{JNxx}
,\\
\mJ_{T,ij}^{xx}
&- k_{T}\xi_{ij,N}^{1/2}(n_{ij}^{y})^{2} + \frac{1}{2}k_{T}\xi_{ij,N}^{-1/2}\xi_{ij,T}n_{ij}^{x}t_{ij}^{x}
,\\
\mJ_{N,ij}^{xy}
&=-k_{N}\left[ \frac{3}{2}+\frac{\xi_{ij,N}}{r_{ij}} \right]\xi_{ij,N}^{1/2}n_{ij}^{x}n_{ij}^{y}
,\\
\mJ_{T,ij}^{xy}
&= k_{T}\xi_{ij,N}^{1/2}n_{ij}^{x}n_{ij}^{y} + \frac{1}{2}k_{T}\xi_{ij,N}^{-1/2}\xi_{ij,T}n_{ij}^{x}t_{ij}^{y}
,\\
\mJ_{N,ij}^{x\ell}
&=0
,\\
\mJ_{T,ij}^{x\ell}
&=k_{T}\xi_{ij,N}^{1/2}n_{ij}^{y}+\frac{1}{2}k_{T}\xi_{ij,N}^{-1/2}\xi_{ij,T}n_{ij}^{x}(n_{ij}^{x}t_{ij}^{y} - n_{ij}^{y}t_{ij}^{x})
,\\
\mJ_{N,ij}^{yx}
&=-k_{N}\left[ \frac{3}{2}+\frac{\xi_{ij,N}}{r_{ij}} \right]\xi_{ij,N}^{1/2}n_{ij}^{x}n_{ij}^{y}
,\\
\mJ_{T,ij}^{yx}
&=k_{T}\xi_{ij,N}^{1/2}n_{ij}^{x}n_{ij}^{y} + \frac{1}{2}k_{T}\xi_{ij,N}^{-1/2}\xi_{ij,T}n_{ij}^{y}t_{ij}^{x}
,\\
\mJ_{N,ij}^{yy}
&=k_{N}\frac{\xi_{ij,N}^{3/2}}{r_{ij}} - k_{N}\left[ \frac{3}{2}+\frac{\xi_{ij,N}}{r_{ij}} \right]\xi_{ij,N}^{1/2}(n_{ij}^{y})^{2}
,\\
\mJ_{T,ij}^{yy}
&=- k_{T}\xi_{ij,N}^{1/2}(n_{ij}^{x})^{2} + \frac{1}{2}k_{T}\xi_{ij,N}^{-1/2}\xi_{ij,T}n_{ij}^{y}t_{ij}^{y}
,\\
\mJ_{N,ij}^{y\ell}
&=0
,\\
\mJ_{T,ij}^{y\ell}
&=-k_{T}\xi_{ij,N}^{1/2}n_{ij}^{x}+\frac{1}{2}k_{T}\xi_{ij,N}^{-1/2}\xi_{ij,T}n_{ij}^{y}(n_{ij}^{x}t_{ij}^{y} - n_{ij}^{y}t_{ij}^{x})
,\\
\mJ_{N,ij}^{\ell x}
&=0
,\\
\mJ_{T,ij}^{\ell x}
&=-k_{T}\xi_{ij,N}^{1/2}n_{ij}^{y}
,\\
\mJ_{N,ij}^{\ell y}
&=0
,\\
\mJ_{T,ij}^{\ell y}
&=k_{T}\xi_{ij,N}^{1/2}n_{ij}^{x}
,\\
\mJ_{N,ij}^{\ell\ell}
&=0
,\\
\mJ_{T,ij}^{\ell \ell}
&=k_{T}\xi_{ij,N}^{1/2}
.
\end{align}
Similarly, the diagonal block elements $\mJ_{N,ij}^{\alpha\beta}, \mJ_{T,ij}^{\alpha\beta}(i=j)$ are given by
\begin{align}
\mJ_{N,ii}^{xx}
&=-\sum_{j\neq i}\left\{
k_{N}\frac{\xi_{ij,N}^{3/2}}{r_{ij}} - k_{N}\left[ \frac{3}{2}+\frac{\xi_{ij,N}}{r_{ij}} \right]\xi_{ij,N}^{1/2}(n_{ij}^{x})^{2}
\right\}
,\\
\mJ_{T,ii}^{xx}
&=-\sum_{j\neq i}\left\{
- k_{T}\xi_{ij,N}^{1/2}(n_{ij}^{y})^{2} + \frac{1}{2}k_{T}\xi_{ij,N}^{-1/2}\xi_{ij,T}n_{ij}^{x}t_{ij}^{x}
\right\}
,\\
\mJ_{N,ii}^{xy}
&=-\sum_{j\neq i}\left\{
-k_{N}\left[ \frac{3}{2}+\frac{\xi_{ij,N}}{r_{ij}} \right]\xi_{ij,N}^{1/2}n_{ij}^{x}n_{ij}^{y}
\right\}
,\\
\mJ_{T,ii}^{xy}
&=-\sum_{j\neq i}\left\{
 k_{T}\xi_{ij,N}^{1/2}n_{ij}^{x}n_{ij}^{y} + \frac{1}{2}k_{T}\xi_{ij,N}^{-1/2}\xi_{ij,T}n_{ij}^{x}t_{ij}^{y}
\right\}
,\\
\mJ_{N,ii}^{x\ell}
&=0
,\\
\mJ_{T,ii}^{x\ell}
&=\sum_{j\neq i}
\left\{
k_{T}\xi_{ij,N}^{1/2}n_{ij}^{y}+\frac{1}{2}\xi_{ij,N}^{-1/2}\xi_{ij,T}n_{ij}^{x}(n_{ij}^{x}t_{ij}^{y} - n_{ij}^{y}t_{ij}^{x})
\right\}
,\\
\mJ_{N,ii}^{yx}
&=-\sum_{j\neq i}\left\{
-k_{N}\left[ \frac{3}{2}+\frac{\xi_{ij,N}}{r_{ij}} \right]\xi_{ij,N}^{1/2}n_{ij}^{x}n_{ij}^{y}
\right\}
,\\
\mJ_{T,ii}^{yx}
&=-\sum_{j\neq i}\left\{
k_{T}\xi_{ij,N}^{1/2}n_{ij}^{x}n_{ij}^{y} + \frac{1}{2}k_{T}\xi_{ij,N}^{-1/2}\xi_{ij,T}n_{ij}^{y}t_{ij}^{x}
\right\}
,\\
\mJ_{N,ii}^{yy}
&=-\sum_{j\neq i}\left\{
k_{N}\frac{\xi_{ij,N}^{3/2}}{r_{ij}} - k_{N}\left[ \frac{3}{2}+\frac{\xi_{ij,N}}{r_{ij}} \right]\xi_{ij,N}^{1/2}(n_{ij}^{y})^{2}
\right\}
,\\
\mJ_{T,ii}^{yy}
&=-\sum_{j\neq i}\left\{
- k_{T}\xi_{ij,N}^{1/2}(n_{ij}^{x})^{2} + \frac{1}{2}k_{T}\xi_{ij,N}^{-1/2}\xi_{ij,T}n_{ij}^{y}t_{ij}^{y}
\right\}
,\\
\mJ_{N,ii}^{yy}
&=0
,\\
\mJ_{T,ii}^{y\ell}
&=-\sum_{j\neq i}
\left\{
k_{T}\xi_{ij,N}^{1/2}n_{ij}^{x}+\frac{1}{2}k_{T}\xi_{ij,N}^{-1/2}\xi_{ij,T}n_{ij}^{y}(n_{ij}^{x}t_{ij}^{y} - n_{ij}^{y}t_{ij}^{x})
\right\}
,\\
\mJ_{N,ii}^{\ell x}
&=0
,\\
\mJ_{T,ii}^{\ell x}
&=\sum_{j\neq i}
k_{T}\xi_{ij,N}^{1/2}n_{ij}^{y}
,\\
\mJ_{N,ii}^{\ell y}
&=0
,\\
\mJ_{T,ii}^{\ell y}
&=-\sum_{j\neq i}
k_{T}\xi_{ij,N}^{1/2}n_{ij}^{x}
,\\
\mJ_{N,ii}^{\ell\ell}
&=0
,\\
\mJ_{T,ii}^{\ell \ell}
&=\sum_{j\neq i}
k_{T}\xi_{ij,N}^{1/2}
\label{JTll}
.
\end{align}
Note that the terms proportional to $\xi_{ij,T}$ in $\mJ_{T}$ include the history-dependent tangential displacements which are ignored in the effective potential \cite{Somfai07,Henkes10,Liu21}.

\section{The detailed derivation of $G$ in the Jacobian analysis \label{AppE}}

In this Appendix, we derive Eq. \eqref{eqEiExp} that gives the rigidity.
First, nonaffine displacements are expanded in terms of eigenfunctions of the Jacobian.
Next, we express the rigidity as the eigenvalues and eigenfunctions of the Jacobian.
Note that we adopt the abbreviation $dA(\bm{q}^{\textrm{FB}}(0))/d\gamma:=dA(\bm{q}(\gamma))/d\gamma|_{\bm{q}(\gamma)=\bm{q}^{\textrm{FB}}(0)}$ in this Appendix.

\subsection{Expansion for nonaffine displacements via eigenfunction of Jacobian \label{AppE1}}
At FB state, $\tilde{F}_i^\alpha/d\gamma$ is expressed as
\begin{align}
\frac{d\tilde F_{i}^{\alpha}}{d\gamma}
&=\lim_{\Delta\gamma\to0}\frac{\tilde F_{i}^{\alpha}(\bm{q}^{\textrm{FB}}(\Delta\gamma))-\tilde F_{i}^{\alpha}(\bm{q}^{\textrm{FB}}(0))}{\Delta\gamma} \nonumber \\
&=\sum_{j\neq i}
\left[
\frac{\partial f_{ij}^{\alpha}}{\partial q_{i}^{x}} y_{ij}(\bm{q}^{\textrm{FB}}(0))
+\sum_{\zeta=x,y} \frac{\partial f_{ij}^{\alpha}}{\partial r_{i}^{\zeta}}\frac{d\mathring r_{ij}^{\zeta}(\bm{q}^{\textrm{FB}}(0))}{d\gamma}
+ \frac{\partial f_{ij}^{\alpha}}{\partial \ell_{i}}  \left(\frac{d\mathring \ell_{i}(\bm{q}^{\textrm{FB}}(0))}{d\gamma} + \frac{d\mathring \ell_{j}(\bm{q}^{\textrm{FB}}(0))}{d\gamma} \right)
\right]
\label{dFdGA}
.
\end{align}
Using the Jacobian, we rewrite Eq. \eqref{dFdGA} as
\begin{align}
\frac{d\tilde F_{i}^{\alpha}}{d\gamma}
&=-\sum_{j\neq i}
\left[
\mJ_{ji}^{\alpha x} y_{ij}(\bm{q}^{\textrm{FB}}(0))
+\sum_{\zeta=x,y} \mJ_{ji}^{\alpha\zeta}\frac{d\mathring r_{ij}^{\zeta}(\bm{q}^{\textrm{FB}}(0))}{d\gamma}
+\mJ_{ji}^{\alpha\ell} \left(\frac{d\mathring \ell_{i}(\bm{q}^{\textrm{FB}}(0))}{d\gamma} + \frac{d\mathring \ell_{j}(\bm{q}^{\textrm{FB}}(0))}{d\gamma} \right)
\right]
\label{Dxy}
,
\end{align}
where the first and second terms on the RHS represent the contributions from the affine and nonaffine displacements, respectively.
Since the affine displacements are applied to our system instantaneously as a step strain, the integral interval of tangential displacements during the affine deformation are zero.
Thus, only the normal contributions in the first term on RHS of Eq. \eqref{Dxy} survive in the affine displacements. 
Then, we rewrite $\mJ_{ij}^{\alpha\beta}$ as $\mJ_{N,ij}^{\alpha\beta}$ in Eq. \eqref{Dxy}: 
\begin{align}
\frac{d\tilde F_{i}^{\alpha}}{d\gamma}
&=-\sum_{j\neq i}
\left[
\mJ_{N,ji}^{\alpha x} y_{ij}(\bm{q}^{\textrm{FB}}(0))
+\sum_{\zeta=x,y} \mJ_{ji}^{\alpha\zeta}\frac{d\mathring r_{ij}^{\zeta}(\bm{q}^{\textrm{FB}}(0))}{d\gamma}
+\mJ_{ji}^{\alpha\ell} \left(\frac{d\mathring \ell_{i}(\bm{q}^{\textrm{FB}}(0))}{d\gamma} + \frac{d\mathring \ell_{j}(\bm{q}^{\textrm{FB}}(0))}{d\gamma} \right)
\right]
\label{Dell}
.
\end{align}

Introducing
\begin{align}
\ket{\Xi_{i}}:=
\left[
\begin{matrix}
&\sum_{j\neq i}\mJ_{N,ji}^{x x}y_{ij} \\
&\sum_{j\neq i}\mJ_{N,ji}^{x y}y_{ij} \\
&\sum_{j\neq i}\mJ_{N,ji}^{x \ell}y_{ij}
\end{matrix}
\right]
\end{align}
and with the aid of $d\tilde F_{i}^{\alpha}/d\gamma=0$ at the FB state in Eq. \eqref{Dell}, 
we obtain
\begin{align}
\Xi_{i}^{\alpha}
&=-\sum_{j\neq i}
\left[
\sum_{\zeta=x,y} \mJ_{ji}^{\alpha\zeta}\frac{d\mathring r_{ij}^{\zeta}}{d\gamma}
+\mJ_{ji}^{\alpha\ell} \left(\frac{d\mathring \ell_{i}}{d\gamma} + \frac{d\mathring \ell_{j}}{d\gamma} \right)
\right]
\label{Xiall}
.
\end{align}

Since $\mJ$ satisfies $\mJ_{ii}^{\kappa\beta}=-\sum_{j\neq i}\mJ_{ji}^{\kappa\beta}$,
we obtain
\begin{align}
\Xi_{i}^{\kappa}
&=
-\sum_{\zeta=x,y}
\left[
\left(\sum_{j\neq i} \mJ_{ji}^{\kappa\zeta}\right)\frac{d\mathring r_{i}^{\zeta}}{d\gamma}
-\sum_{j\neq i}\mJ_{ji}^{\kappa\zeta}\frac{d\mathring r_{j}^{\zeta}}{d\gamma}
\right]
-\left[
\left(\sum_{j\neq i}\mJ_{ji}^{\kappa\ell}\right) \frac{d\mathring \ell_{i}}{d\gamma}
+\sum_{j\neq i}\mJ_{ji}^{\kappa\ell} \frac{d\mathring \ell_{j}}{d\gamma}
\right]
\nonumber \\
&=
-\sum_{\zeta=x,y}
\left[
-\mJ_{ii}^{\kappa\zeta}\frac{d\mathring r_{i}^{\zeta}}{d\gamma}
-\sum_{j\neq i}\mJ_{ji}^{\kappa\zeta}\frac{d\mathring r_{j}^{\zeta}}{d\gamma}
\right]
-\left[
-\mJ_{ii}^{\kappa\ell}\frac{d\mathring \ell_{i}}{d\gamma}
+ \sum_{j\neq i}\mJ_{ji}^{\kappa\ell} \frac{d\mathring \ell_{j}}{d\gamma}
\right]
\nonumber \\
&=
\sum_{\zeta=x,y}
\sum_{j=1}^{N}\mJ_{ji}^{\kappa\zeta}\frac{d\mathring r_{j}^{\zeta}}{d\gamma}
+ \mJ_{ii}^{\kappa\ell}\frac{d\mathring \ell_{i}}{d\gamma} - \sum_{j\neq i}\mJ_{ji}^{\kappa\ell} \frac{d\mathring \ell_{j}}{d\gamma}
\label{XiHessian}
.
\end{align}
Since $\mJ$ satisfies $\mJ_{ii}^{\ell\beta}=\sum_{j\neq i}\mJ_{ji}^{\ell\beta}$, we obtain
\begin{align}
\Xi_{i}^{\ell}
&=-
\sum_{\zeta=x,y}
\left[
\left(\sum_{j\neq i}\mJ_{ji}^{\ell\zeta}\right) \frac{d\mathring r_{i}^{\zeta}}{d\gamma}
-\sum_{j\neq i}\mJ_{ji}^{\ell\zeta} \frac{d\mathring r_{j}^{\zeta}}{d\gamma}
\right]
-
\left[
\left(\sum_{j\neq i}\mJ_{ji}^{\ell\ell}\right)\frac{d\mathring \ell_{i}}{d\gamma}
+\sum_{j\neq i}\mJ_{ji}^{\ell\ell}\frac{d\mathring \ell_{j}}{d\gamma}
\right]
\nonumber \\
&=-
\sum_{\zeta=x,y}
\left[
\mJ_{ii}^{\ell\zeta} \frac{d\mathring r_{i}^{\zeta}}{d\gamma}
-\sum_{j\neq i}\mJ_{ji}^{\ell\zeta} \frac{d\mathring r_{j}^{\zeta}}{d\gamma}
\right]
+
\left[
\mJ_{ii}^{\ell\ell}\frac{d\mathring \ell_{i}}{d\gamma}
+\sum_{j\neq i}\mJ_{ji}^{\ell\ell}\frac{d\mathring \ell_{j}}{d\gamma}
\right]
\nonumber \\
&=
-\sum_{\zeta=x,y}
\left[
\mJ_{ii}^{\ell\zeta} \frac{d\mathring r_{i}^{\zeta}}{d\gamma}
-\sum_{j\neq i}\mJ_{ji}^{\ell\zeta} \frac{d\mathring r_{j}^{\zeta}}{d\gamma}
\right]
+
\sum_{j=1}^{N}\mJ_{ji}^{\ell\ell}\frac{d\mathring \ell_{j}}{d\gamma}
\label{XiHessianEll}
.
\end{align}
Let us introduce $\tilde\mJ_{ii}^{\alpha\beta}$ as
\begin{align}
\tilde\mJ_{ii}^{\alpha\beta}
&:=
\left\{
\begin{matrix}
-\mJ_{ii}^{\ell x} & (\alpha=\ell,\ \beta=x) \\
-\mJ_{ii}^{\ell y} & (\alpha=\ell,\ \beta=y) \\
\mJ_{ii}^{\alpha\beta} & (\textrm{otherwise})
\end{matrix}
\right.
\label{tildeJ3}
\end{align}
and
\begin{align}
\tilde\mJ_{ij}^{\alpha\beta}:=\mJ_{ij}^{\alpha\beta}
.
\end{align}
Here, $\tilde\mJ_{ij}$ satisfies
\begin{align}
\tilde\mJ_{ji}^{\alpha\beta}
&=
\left\{
\begin{matrix}
-\mJ_{ij}^{x\ell} & (\alpha=x,\ \beta=\ell) \\
-\mJ_{ij}^{y\ell} & (\alpha=y,\ \beta=\ell) \\
\mJ_{ij}^{\alpha\beta} & (\textrm{otherwise}) .
\end{matrix}
\right.
\label{tildeJ}
\end{align}
With the aid of $\tilde\mJ$ Eqs. \eqref{XiHessian} and \eqref{XiHessianEll} are rewritten as
\begin{align}
\Xi_{i}^{\alpha}
&=
\sum_{\zeta=x,y}
\sum_{j=1}^{N}\tilde\mJ_{ij}^{\alpha\zeta}\frac{d\mathring r_{j}^{\zeta}}{d\gamma}
+ \tilde\mJ_{ii}^{\alpha\ell}\frac{d\mathring \ell_{i}}{d\gamma} + \sum_{j\neq i}\tilde\mJ_{ij}^{\alpha\ell} \frac{d\mathring \ell_{j}}{d\gamma}
\nonumber \\
&=
\sum_{\zeta=x,y}
\sum_{j=1}^{N}\tilde\mJ_{ij}^{\alpha\zeta}\frac{d\mathring r_{j}^{\zeta}}{d\gamma}
+\sum_{j=1}^{N}\tilde\mJ_{ij}^{\alpha\ell} \frac{d\mathring \ell_{j}}{d\gamma}
\nonumber \\
&=
\sum_{\beta=x,y,\ell}
\sum_{j=1}^{N}\tilde\mJ_{ij}^{\alpha\beta}\frac{d\mathring q_{j}^{\beta}}{d\gamma}
,
\label{XiXY}
\\
\Xi_{i}^{\ell}
&=
\sum_{\zeta=x,y}
\left[
\mJ_{ii}^{\ell\zeta} \frac{d\mathring r_{i}^{\zeta}}{d\gamma}
+\sum_{j\neq i}\mJ_{ij}^{\ell\zeta} \frac{d\mathring r_{j}^{\zeta}}{d\gamma}
\right]
+
\sum_{j=1}^{N}\mJ_{ji}^{\ell\ell}\frac{d\mathring \ell_{j}}{d\gamma}
\nonumber \\
&=
\sum_{\zeta=x,y}
\sum_{j=1}^{N}\mJ_{ij}^{\ell\zeta} \frac{d\mathring r_{j}^{\zeta}}{d\gamma}
+
\sum_{j=1}^{N}\mJ_{ji}^{\ell\ell}\frac{d\mathring \ell_{j}}{d\gamma}
\nonumber \\
&=
\sum_{\beta=x,y,\ell}
\sum_{j=1}^{N}\mJ_{ij}^{\ell\beta} \frac{d\mathring q_{j}^{\beta}}{d\gamma}
\label{XiEll}
.
\end{align}
Equations \eqref{XiXY} and \eqref{XiEll} can be rewritten as
\begin{align}
\Xi_{i}^{\alpha}
&=
\sum_{j=1}^{N}
\sum_{\beta=x,y,\ell} \tilde\mJ_{ij}^{\alpha\beta}\frac{d\mathring q_{j}^{\beta}}{d\gamma}
\label{Base}
.
\end{align}
Furthermore, Eq. \eqref{Base} can be expressed as
\begin{align}
\Ket{\Xi}
=\tilde\mJ\Ket{\frac{d\mathring q}{d\gamma}},
\label{base}
\end{align}
which corresponds to Eq. \eqref{dFdG2} in the main text,
where $\Ket{d\mathring q/d\gamma}$ is introduced in Eq. \eqref{NonAff}.

Let us expand $\ket{d\mathring q/d\gamma}$ by the right eigenfunction $\ket{\tilde R_{n}}$ of $\tilde\mJ$ as
\begin{align}
\Ket{\frac{d\mathring q}{d\gamma}}
&=a_{n}\ket{\tilde R_{n}}
.
\label{exp}
\end{align}
Substituting Eq. \eqref{exp} into Eq. \eqref{base}, we obtain
\begin{align}
\Ket{\Xi}
&= \tilde\lambda_{n}a_{n}\ket{\tilde R_{n}}
.
\label{D16}
\end{align}
Multiplying $\bra{\tilde L_{m}}$ to Eq. \eqref{D16} with the aid of the orthonormal relation,
we obtain
\begin{align}
\Braket{\tilde L_{m}|\Xi}
&=\tilde\lambda_{n}a_{n}\Braket{\tilde L_{m}|\tilde R_{n}} \nonumber \\
&=\tilde\lambda_{m}a_{m}.
\end{align}
Substituting this into Eq. \eqref{exp}, we obtain Eq. \eqref{eqEiExp}.

\subsection{The expression of $G$ \label{AppC2}}
Let us evaluate the rigidity $G$ defined as Eq. \eqref{Gdef}.
Substituting Eqs. \eqref{Sigma} and \eqref{g} into Eq. \eqref{Gdef}, we obtain
\begin{align}
G
&=-\Braket{
\lim_{\Delta\gamma\to0}\frac{1}{2\Delta\gamma L^{2}}\sum_{i,j (i\neq j)} \left[ f_{ij}^{x}(\bm{q}^{\textrm{FB}}(\Delta\gamma))r_{ij}^{y}(\bm{q}^{\textrm{FB}}(\Delta\gamma)) - f_{ij}^{x}(\bm{q}^{\textrm{FB}}(0))r_{ij}^{y}(\bm{q}^{\textrm{FB}}(0)) \right]
}
\label{defG}
,
\end{align}
where we have adopted the symmetric expression for $i$ and $j$ in the summation in Eq. \eqref{defG}.

Expanding $r_{ij}^{\alpha}(\bm{q}^{\textrm{FB}}(\Delta\gamma))$ in Eq. \eqref{defG} by $\Delta\gamma$ from the zero strain state, we obtain
\begin{align}
r_{ij}^{\alpha}(\bm{q}^{\textrm{FB}}(\Delta\gamma))
&=r_{i}^{\alpha}(\bm{q}^{\textrm{FB}}(\Delta\gamma))-r_{j}^{\alpha}(\bm{q}^{\textrm{FB}}(\Delta\gamma)) \nonumber \\
&\simeq r_{ij}^{\alpha}(\bm{q}^{\textrm{FB}}(0)) + \Delta\gamma\left\{ \delta_{\alpha x}\left(y_{i}(\bm{q}^{\textrm{FB}}(0))-y_{j}(\bm{q}^{\textrm{FB}}(0))\right) + \frac{d\mathring r_{i}^{\alpha}(\bm{q}^{\textrm{FB}}(0))}{d\gamma} - \frac{d\mathring r_{j}^{\alpha}(\bm{q}^{\textrm{FB}}(0))}{d\gamma} \right\} \nonumber \\
&=r_{ij}^{\alpha}(\bm{q}^{\textrm{FB}}(0)) + \Delta\gamma\left\{ \delta_{\alpha x}y_{ij}(\bm{q}^{\textrm{FB}}(0))+\frac{d\mathring r_{ij}^{\alpha}(\bm{q}^{\textrm{FB}}(0))}{d\gamma} \right\}
\label{rij}
.
\end{align}

Similarly, expanding $f_{ij}^{\alpha}(\Delta\gamma)$ in Eq. \eqref{defG} from the zero strain state, we obtain
\begin{align}
f_{ij}^{\alpha}(\bm{q}^{\textrm{FB}}(\Delta\gamma))
&\simeq f_{ij}^{\alpha}(\bm{q}^{\textrm{FB}}(0))
+ \sum_{k=1}^{N}\sum_{\zeta=x,y}\Delta\gamma\frac{\partial f_{ij}^{\alpha}}{\partial r_{k}^{\zeta}}\frac{d r_{k}^{\zeta}}{d\gamma}
+ \sum_{k=1}^{N}\Delta\gamma\frac{\partial f_{ij}^{\alpha}}{\partial \ell_{k}}\frac{d \ell_{k}}{d\gamma}
 \nonumber \\
&=f_{ij}^{\alpha}(\bm{q}^{\textrm{FB}}(0))
+ \sum_{\zeta=x,y}\Delta\gamma\left[
\frac{\partial f_{ij}^{\alpha}}{\partial r_{i}^{\zeta}}\Biggl( \delta_{\zeta x}y_{i}(\bm{q}^{\textrm{FB}}(0))+\frac{d\mathring r_{i}^{\zeta}(\bm{q}^{\textrm{FB}}(0))}{d\gamma} \right)
+
\frac{\partial f_{ij}^{\alpha}}{\partial r_{j}^{\zeta}}\left( \delta_{\zeta x}y_{j}(\bm{q}^{\textrm{FB}}(0))+\frac{d\mathring r_{j}^{\zeta}(\bm{q}^{\textrm{FB}}(0))}{d\gamma} \Biggl)
\right]
\nonumber \\
&\quad
+ \Delta\gamma\left[
\frac{\partial f_{ij}^{\alpha}}{\partial \ell_{i}}\Biggl( \delta_{\ell x}y_{i}(\bm{q}^{\textrm{FB}}(0))+\frac{d\mathring \ell_{i}(\bm{q}^{\textrm{FB}}(0))}{d\gamma} \right)
+
\frac{\partial f_{ij}^{\alpha}}{\partial \ell_{j}}\left( \delta_{\ell x}y_{j}(\bm{q}^{\textrm{FB}}(0))+\frac{d\mathring \ell_{j}(\bm{q}^{\textrm{FB}}(0))}{d\gamma} \Biggl)
\right]
.
\end{align}
Furthermore, using ${\partial f_{ij}^{\alpha}}/{\partial r_{j}^{\zeta}}=-{\partial f_{ij}^{\alpha}}/{\partial r_{i}^{\zeta}}$ and $ {\partial f_{ij}^{\alpha}}/{\partial \ell_{j}}={\partial f_{ij}^{\alpha}}/{\partial \ell_{i}}$,
$f_{ij}^{\alpha}$ can be written as
\begin{align}
f_{ij}^{\alpha}(\bm{q}^{\textrm{FB}}(\Delta\gamma))
&=f_{ij}^{\alpha}(\bm{q}^{\textrm{FB}}(0))
+\sum_{\zeta=x,y} \Delta\gamma \frac{\partial f_{ij}^{\alpha}}{\partial r_{i}^{\zeta}}\left( \delta_{\zeta x}y_{ij}(\bm{q}^{\textrm{FB}}(0))+\frac{d\mathring r_{ij}^{\zeta}(\bm{q}^{\textrm{FB}}(0))}{d\gamma} \right)
+ \Delta\gamma \frac{\partial f_{ij}^{\alpha}}{\partial \ell_{i}}  \left(\frac{d\mathring \ell_{i}(\bm{q}^{\textrm{FB}}(0))}{d\gamma} + \frac{d\mathring \ell_{j}(\bm{q}^{\textrm{FB}}(0))}{d\gamma} \right)
\label{fij}
.
\end{align}

Substituting Eqs. \eqref{rij} and \eqref{fij} into Eq. \eqref{defG}, we obtain
\begin{align}
G
&=
-\frac{1}{2 L^{2}}
\left\langle
\sum_{i,j(i\neq j)}
\Biggl[
f_{ij}^{x}(\bm{q}^{\textrm{FB}}(0))\frac{d\mathring q_{ij}^{y}(\bm{q}^{\textrm{FB}}(0))}{d \gamma}
+\sum_{\zeta=x,y}\frac{\partial f_{ij}^{x}(\bm{q}^{\textrm{FB}}(0))}{\partial r_{i}^{\zeta}}r_{ij}^{y}(\bm{q}^{\textrm{FB}}(0))
\left( \delta_{\zeta x}y_{ij}(\bm{q}^{\textrm{FB}}(0))+\frac{d\mathring r_{ij}^{\zeta}(\bm{q}^{\textrm{FB}}(0))}{d\gamma} \right)
\right.
\nonumber \\
&\quad\quad\quad\quad\quad\quad\quad\quad\quad\quad\quad\quad
+
\left.
\frac{\partial f_{ij}^{x}(\bm{q}^{\textrm{FB}}(0))}{\partial \ell_{i}}r_{ij}^{y}(\bm{q}^{\textrm{FB}}(0))
 \left(\frac{d\mathring \ell_{i}(\bm{q}^{\textrm{FB}}(0))}{d\gamma} + \frac{d\mathring \ell_{j}(\bm{q}^{\textrm{FB}}(0))}{d\gamma}\right)
\Biggl]
\right\rangle
\label{eq52}
.
\end{align}
Because $\sum_{i(i\neq j)}f_{ij}^{\alpha}(\bm{q}^{\textrm{FB}}(0))=0$ at the FB state, the first term on RHS of Eq. \eqref{eq52} can be written as
\begin{align}
\sum_{i,j(i\neq j)}f_{ij}^{x}(\bm{q}^{\textrm{FB}}(0))\frac{d\mathring{q}_{ij}^{y}(\bm{q}^{\textrm{FB}}(0))}{d\gamma}
&=\sum_{i,j(i\neq j)}f_{ij}^{x}(\bm{q}^{\textrm{FB}}(0))\left( \frac{d\mathring{q}_{i}^{y}(\bm{q}^{\textrm{FB}}(0))}{d\gamma} - \frac{d\mathring{q}_{j}^{y}(\bm{q}^{\textrm{FB}}(0))}{d\gamma}\right)
\nonumber \\
&=\sum_{j}\left(\sum_{j(j\neq i)}f_{ij}^{x}(\bm{q}^{\textrm{FB}}(0))\right) \frac{d\mathring{q}_{i}^{y}(\bm{q}^{\textrm{FB}}(0))}{d\gamma}
-\sum_{i}\left(\sum_{i(i\neq j)} f_{ij}^{x}(\bm{q}^{\textrm{FB}}(0))\right) \frac{d\mathring{q}_{j}^{y}(\bm{q}^{\textrm{FB}}(0))}{d\gamma}
\nonumber \\
&=0
\label{eq53}
.
\end{align}
Thus, $G$ is expressed as
 \begin{align}
 G
&=-
\frac{1}{2 L^{2}}
\left\langle
\sum_{i,j(i\neq j)}
\left[
\sum_{\zeta=x,y}\frac{\partial f_{ij}^{x}(\bm{q}^{\textrm{FB}}(0))}{\partial r_{i}^{\zeta}}y_{ij}(\bm{q}^{\textrm{FB}}(0))\left( \delta_{\zeta x}y_{ij}(\bm{q}^{\textrm{FB}}(0))+\frac{d\mathring r_{ij}^{\zeta}(\bm{q}^{\textrm{FB}}(0))}{d\gamma} \right)
\right.
\right.
\nonumber \\
&\quad\quad\quad\quad\quad\quad\quad\quad
\left.
\left.
+\frac{\partial f_{ij}^{x}(\bm{q}^{\textrm{FB}}(0))}{\partial \ell_{i}}y_{ij}(\bm{q}^{\textrm{FB}}(0))
 \left(\frac{d\mathring \ell_{i}(\bm{q}^{\textrm{FB}}(0))}{d\gamma} + \frac{d\mathring \ell_{j}(\bm{q}^{\textrm{FB}}(0))}{d\gamma} \right)
\right]
\right \rangle
.
\end{align}
With the aid of $\mJ_{ij}^{\alpha\beta}:=-\partial_{q_{j}^{\beta}}f_{ij}^{\alpha}\ (i\neq j)$, we can express $G$ as
\begin{align}
G
&=
\frac{1}{2 L^{2}}
\left\langle
\sum_{i,j(i\neq j)}
\left[
\sum_{\zeta=x,y}y_{ij}(\bm{q}^{\textrm{FB}}(0))\mJ_{ji}^{x\zeta}(\bm{q}^{\textrm{FB}}(0))\left( \delta_{\zeta x}y_{ij}(\bm{q}^{\textrm{FB}}(0))+\frac{d\mathring r_{ij}^{\zeta}(\bm{q}^{\textrm{FB}}(0))}{d\gamma} \right)
\right.
\right.
\nonumber \\
&\quad\quad\quad\quad\quad\quad\quad
\left.
\left.
+y_{ij}(\bm{q}^{\textrm{FB}}(0))\mJ_{ji}^{x\ell}(\bm{q}^{\textrm{FB}}(0))\left( \frac{d\mathring \ell_{i}(\bm{q}^{\textrm{FB}}(0))}{d\gamma} + \frac{d\mathring \ell_{j}(\bm{q}^{\textrm{FB}}(0))}{d\gamma} \right)
\right]
\right\rangle
\nonumber \\
&=
\frac{1}{2 L^{2}}
\left\langle
\sum_{i,j(i\neq j)}
\left[
y_{ij}^{2}(\bm{q}^{\textrm{FB}}(0))\mJ_{ji}^{xx}(\bm{q}^{\textrm{FB}}(0))
+\sum_{\zeta=x,y}y_{ij}(\bm{q}^{\textrm{FB}}(0))\mJ_{ji}^{x\zeta}(\bm{q}^{\textrm{FB}}(0))\frac{d\mathring r_{ij}^{\zeta}(\bm{q}^{\textrm{FB}}(0))}{d\gamma}
\right.
\right.
\nonumber \\
&\quad\quad\quad\quad\quad\quad\quad
\left.
\left.
+y_{ij}(\bm{q}^{\textrm{FB}}(0))\mJ_{ji}^{x\ell}(\bm{q}^{\textrm{FB}}(0))\left( \frac{d\mathring \ell_{i}(\bm{q}^{\textrm{FB}}(0))}{d\gamma} + \frac{d\mathring \ell_{j}(\bm{q}^{\textrm{FB}}(0))}{d\gamma} \right)
\right]
\right\rangle
\label{eqG}
.
\end{align}
Thus, with Eqs \eqref{tildeJ1} and \eqref{tildeJ2},
we obtain Eqs. \eqref{expG}-\eqref{GNAff}.

\section{DOS in terms of the effective Hessian \label{AppF}}

In this Appendix, we introduce the DOS with the aid of the effective Hessian as in Refs. \cite{Somfai07,Henkes10,Liu21}.
The effective Hessian $\mH$ at the FB state is defined as
\begin{align}
\mH_{ij}^{\alpha\beta}:=\left. \frac{\partial^{2}U_{\textrm{eff}}}{\partial q_{i}^{\alpha} \partial q_{j}^{\beta}} \right|_{\bm{q}(\gamma)=\bm{q}^{\textrm{FB}}(0)},
\end{align}
where $U_{\textrm{eff}}$ is the effective potential defined as
\begin{align}
U_{\textrm{eff}}
:=\frac{1}{2}\sum_{\braket{ij}}
\left[ k_{N}(\delta\bm{r}_{ij}\cdot\bm{n}_{ij})^{2} - \frac{|\bm{f}_{N,ij}|}{r_{ij}^{\textrm{FB}}}(\delta\bm{r}_{ij}\cdot\bm{t}_{ij})^{2} + k_{T}\delta t_{ij}^{2} \right]
\end{align}
with $\delta\bm{r}_{ij}:=\delta\bm{r}_{i}-\delta\bm{r}_{j}, \delta\bm{r}_{i}:=\bm{r}_{i}-\bm{r}_{i}^{\textrm{FB}}, r_{ij}^{\textrm{FB}}:=|\bm{r}^{\textrm{FB}}_{i}-\bm{r}_{j}^{\textrm{FB}}|, \delta t_{ij}:=\delta\bm{r}_{ij}\cdot\bm{t}_{ij}-(\delta\ell_{i}+\delta\ell_{j})$, and $\delta\ell_{i}:=\ell_{i}-\ell_{i}^{\textrm{FB}}$.
Here, $\bm{r}_{i}^{\textrm{FB}}$ and $\ell_{i}^{\textrm{FB}}$ are the position of $i$-th particle and $3$rd component of $\bm{q}_{i}$ at the FB state, respectively.
Thus, $\mH$ is a $3N\times3N$ matrix corresponding to the Jacobian.
We note that this Hessian matrix is a real symmetric matrix, and thus, it can be diagonalized by an orthogonal matrix,
where the eigenvectors are orthogonal with each other, and the corresponding eigenvalues are real number.

The eigenvalue equation of $\mH$ is expressed as
\begin{align}
\mH\ket{n}=\lambda_{H,n}\ket{n},
\end{align}
where $\lambda_{H,n}$ and $\ket{n}$ are the $n$-th eigenvalue and eigenvector of $\mH$, respectively.
Note that the left eigenvalue is also given by $\langle n|\mathcal{H}=\lambda_{H,n}\langle n|$,
 where $\langle n|=|n\rangle^T$.
Then, we introduce the DOS $D_{H}$ in terms of $\mH$ as
\begin{align}
D_{H}(\omega):=\frac{1}{3N}\sum_{n}\nolimits' \langle \delta(\omega-\omega_{H,n})\rangle
,
\end{align}
where $\omega_{H,n}:=\sqrt{\lambda_{H,n}}$.

\section{System size dependence for the DOS \label{SystemSize}}

The system size dependence of the DOS is investigated in this Appendix.
From Fig. \ref{figDOSSystemSize} we have confirmed that both the rotational and translational bands show little dependence on system size. 
This means that the rotational band is not a virtual band that can only be observed in a small system, but an intrinsic band that can be observed in the thermodynamic limit.
Thus, we expect that our observed results will remain unchanged even if we are interested in larger systems.

\begin{figure}[htbp]
\centering
\includegraphics[width=9cm]{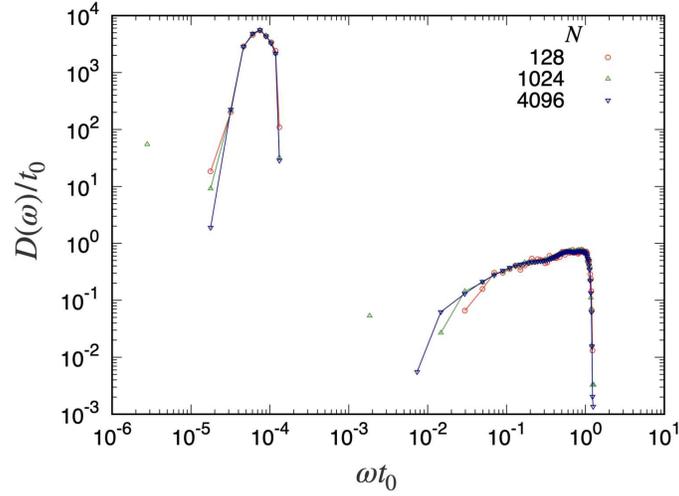}
\caption{
Double logarithmic plots of $D(\omega)$ against $\omega t_0$ for various $N$ at $k_{T}/k_{N}=1.0\times10^{-8}$, $\phi=0.90$. 
}
\label{figDOSSystemSize}
\end{figure}

\section{Density dependence for DOS \label{DensityDep}}

In this Appendix, we investigate the density dependence for the DOS.
As shown in Fig. \ref{figDOSDensity} for $k_T/k_N=10^{-8}$, the DOS depends on $\phi$, where the rotational band shifts to a lower frequency region and the plateau of the translational band becomes longer as the density approaches the jamming point. 
The latter result is well known from previous studies such as Ref.~\cite{Wyart05}.

\begin{figure}[htbp]
\centering
\includegraphics[width=9cm]{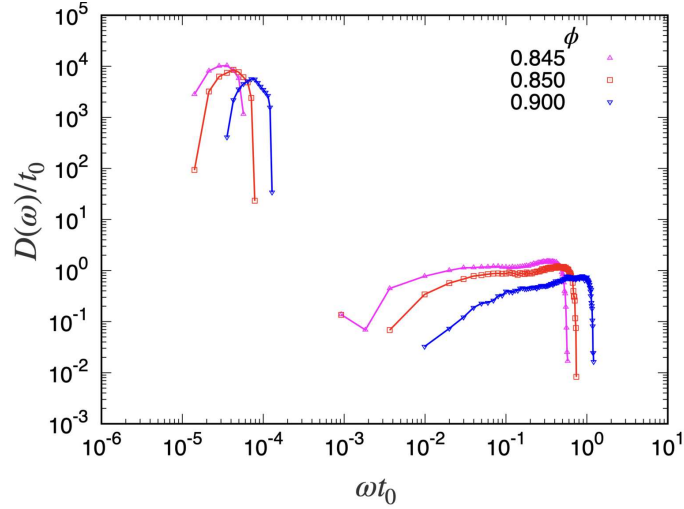}
\caption{
Double logarithmic plots of $D(\omega)$ against $\omega t_0$ for various $\phi$ at $k_T/k_N=1.0\times10^{-8}$ and $N=4096$. 
}
\label{figDOSDensity}
\end{figure}

\end{widetext}


\end{document}